\documentclass{article}
\usepackage[utf8]{inputenc}
\usepackage{fullpage}
\usepackage{float}
\usepackage{amsmath,amsthm, amsfonts}
\usepackage{graphicx,caption}
\usepackage{xcolor}
\usepackage{authblk}
\usepackage{arydshln}

\DeclareMathOperator*{\argmin}{arg\,min}
\newcommand{\Var}{\mathrm{Var}}

\newcommand{\keywords}[1]{\textbf{\textit{Keywords---}} #1}

\usepackage[citestyle = numeric, bibstyle =  apa, sorting = none]{biblatex}
\makeatletter
\RequireBibliographyStyle{numeric}

\title{Standard error estimation in meta-analysis of studies reporting medians}
\author[1]{Sean McGrath}
\author[2]{Stephan Katzenschlager}
\author[3,4]{Alexandra J.\ Zimmer}
\author[5]{Alexander Seitel}
\author[6]{Russell Steele}
\author[3,7,8]{Andrea Benedetti}
\date{}

\affil[1]{\small Department of Biostatistics, Harvard T.H. Chan School of Public Health, Boston, MA, USA}
\affil[2]{\small Department of Anesthesiology, Heidelberg University Hospital, Heidelberg, Germany}
\affil[3]{\small Department of Epidemiology, Biostatistics, and Occupational Health, McGill University, Montreal, Quebec, Canada}
\affil[4]{\small McGill International TB Centre, Montreal, Quebec, Canada}
\affil[5]{\small Division of Intelligent Medical Systems, German Cancer Research Center (DKFZ), Heidelberg, Germany}
\affil[6]{\small Department of Mathematics and Statistics, McGill University, Montreal, Quebec, Canada}
\affil[7]{\small Respiratory Epidemiology and Clinical Research Unit (RECRU), McGill University Health Centre, Montreal, Quebec, Canada}
\affil[8]{\small Department of Medicine, McGill University, Montreal, Quebec, Canada}

\begin{document}

\maketitle

\begin{abstract}
    We consider the setting of an aggregate data meta-analysis of a continuous outcome of interest. When the distribution of the outcome is skewed, it is often the case that some primary studies report the sample mean and standard deviation of the outcome and other studies report the sample median along with the first and third quartiles and/or minimum and maximum values. To perform meta-analysis in this context, a number of approaches have recently been developed to impute the sample mean and standard deviation from studies reporting medians. Then, standard meta-analytic approaches with inverse-variance weighting are applied based on the (imputed) study-specific sample means and standard deviations. In this paper, we illustrate how this common practice can severely underestimate the within-study standard errors, which results in overestimation of between-study heterogeneity in random effects meta-analyses. We propose a straightforward bootstrap approach to estimate the standard errors of the imputed sample means. Our simulation study illustrates how the proposed approach can improve estimation of the within-study standard errors and between-study heterogeneity. Moreover, we apply the proposed approach in a meta-analysis to identify risk factors of a severe course of COVID-19.

\end{abstract}


\section{Introduction}

Meta-analysis is a statistical approach that synthesizes data across studies addressing a common research question. We consider the setting where the outcome of interest in a meta-analysis is continuous. In this setting, primary studies typically report the sample mean and standard deviation of the outcome. However, when the distribution of the outcome is skewed, primary studies often instead report the sample median of the outcome along with the first and third quartiles and/or the minimum and maximum values \cite{higgins2020cochrane}. 

A number of approaches have recently been developed to perform meta-analysis when some primary studies report sample means and others report sample medians. The most commonly applied approach is to first impute the sample mean and standard deviation from studies that report medians along with the first and third quartiles and/or minimum and maximum values. Then, one applies standard meta-analytic approaches with inverse-variance weighting based on the (imputed) study-specific sample means and standard deviations. Such approaches, which we refer to as \emph{transformation-based approaches}, were first proposed and systematically evaluated by Hozo et al.\ \cite{hozo2005estimating} and have been further developed by a number of authors in recent years \cite{bland2015estimating, wan2014estimating, kwon2015simulation, luo2018optimally, mcgrath2020estimating, shi2020optimally, shi2020estimating, walter2022estimation, cai2021estimating, yang2021generalized}. Reflecting their widespread application, Google Scholar lists over 10,000 articles citing these transformation-based approaches (i.e., \cite{hozo2005estimating, bland2015estimating, wan2014estimating, kwon2015simulation, luo2018optimally, mcgrath2020estimating, shi2020optimally, shi2020estimating, walter2022estimation, cai2021estimating, yang2021generalized}) as of May 1 2022.

A fundamental problem of the standard application of transformation-based approaches is that the sampling variability of the imputed sample means is not appropriately accounted for when performing an inverse-variance weighted meta-analysis. For instance, consider the simple case of a meta-analysis of one-group studies that report the sample median of the outcome. While the standard deviation divided by the square root of the sample size is the standard error (SE) of the sample mean, this is not necessarily equal to the SE of the mean estimator of the transformation-based approaches. In particular, such mean estimators are typically much less efficient than the sample mean because they are based on less data. Consequently, the estimated standard deviation divided by the square root of the sample size -- which we refer to as the \emph{naïve SE estimator} -- may underestimate the SE of the mean estimator for studies reporting medians. 

Underestimation of within-study SEs can have several downstream consequences in meta-analysis, which we investigate in this paper. For instance, consider the between-study variance estimator in a random effects meta-analysis. Since the total variance of the study-specific means decomposes into the sum of the within-study variance and between-study variance, systematically underestimating the within-study SEs results in overestimating the between-study variance. This in turn introduces bias in estimating other parameters that depend on the between-study variance, especially those quantifying between-study heterogeneity such as the $I^2$ index \cite{higgins2002quantifying, higgins2003measuring}. More generally, poor estimation of between-study heterogeneity may strongly affect the conclusions meta-analyses, as the degree of between-study heterogeneity helps researchers understand the generalizability of their conclusions, helps guide additional analyses to identify sources of heterogeneity (e.g., subgroup analyses, meta-regression analyses), and even helps determine whether it makes sense to pool the data in a meta-analysis at all \cite{higgins2020cochrane}.

The use of the naïve SE estimators in the literature presumably arose because most of the papers proposing transformation-based approaches suggested that data analysts use the imputed study-specific sample means and standard deviations in place of the actual, unreported sample means and standard deviations in standard meta-analytic approaches. Even more explicitly, Hozo et al.\ \cite{hozo2005estimating}, Luo et al.\ \cite{luo2018optimally}, McGrath et al.\ \cite{mcgrath2020estimating}, and Shi et al.\ \cite{shi2020estimating} used the the naïve SE estimator in data applications illustrating how to apply their proposed transformation-based approaches, for instance. While some have raised concerns over potential issues of the naïve SE estimator (e.g., see the open peer review report of Hozo et al.\ \cite{hozo2005estimating}), the downstream consequences of using naïve SE estimators has been largely unexplored in the literature because most simulation studies evaluating the transformation-based approaches (e.g., see \cite{hozo2005estimating, bland2015estimating, wan2014estimating, kwon2015simulation, luo2018optimally, mcgrath2020estimating, shi2020optimally, shi2020estimating, walter2022estimation, cai2021estimating}) have almost entirely focused on the performance of these methods at the study-level (i.e., for estimating the mean and standard deviation of the outcome distribution for a given study) rather than at the meta-analytic level. 

Analytically deriving the SE of the mean estimator of the transformation-based approaches is case-specific and often challenging, as several of the currently best-performing approaches involve model selection and are solutions to ad-hoc optimization problems. However, since most of these transformation-based approaches assume parametric models, parametric bootstrap \cite{efron1994introduction} is a natural approach to consider for estimating the SE of the mean estimator, which we explore in this paper.

The methodological contributions of this paper are as follows: 1) Focusing on the recently proposed transformation-based approaches of McGrath et al.\ \cite{mcgrath2020estimating} and Cai et al.\ \cite{cai2021estimating}, we illustrate that the standard application of transformation-based approaches severely underestimates the within-study SEs and consequently overestimates between-study heterogeneity in random effects meta-analyses. 2) We describe a parametric bootstrap approach to estimate within-study SEs in this context and systematically evaluate its performance at both the study- and meta-analytic level. 

In Section \ref{sec:methods}, we describe the transformation-based approaches and the proposed bootstrap SE estimator. In Section \ref{sec:simulations}, we perform a simulation study evaluating the performance of the naïve and bootstrap SE estimators at both the study- and meta-analytic level. We perform an empirical comparison of the naïve and bootstrap SE estimators in a meta-analysis of risk factors of a severe course of COVID-19 in Section \ref{sec:application}. We conclude with a discussion in Section \ref{sec:discussion}.

\section{Methods} \label{sec:methods}

\subsection{Meta-analytic models, estimands, and estimators} \label{sec:models}

Consider an aggregate data meta-analysis of a continuous outcome of interest. For each primary study $k$ ($k = 1, \dots, K$), let $y_k$ denote the point estimate of the outcome of interest $\sigma_k^2$ denote the estimate of its sampling variance. For instance, $y_k$ may represent the sample mean of the outcome of interest. The study-specific estimates are assumed to be distributed as
\begin{equation*}
    y_k \sim \text{Normal}(\mu_k, \sigma_k^2), \qquad k = 1, \dots, K
\end{equation*}
where the $\sigma_k^2$ are regarded as known quantities. The \emph{common effect} (also referred to as \emph{fixed effect}) model assumes that the study-specific means are identical (i.e., $\mu_k = \mu_0$ for $k = 1, \dots, K$, $\mu_0 \in \mathbb{R}$). The \emph{random effects} model assumes that the study-specific means are random variables that are distributed as
\begin{equation*}
    \mu_k \sim \text{Normal}(\mu_{\text{pool}}, \tau^2), \qquad k = 1, \dots, K.
\end{equation*}
We refer to $\mu_{\text{pool}}$ as the \emph{pooled mean} and $\tau^2$ as the \emph{between-study variance}. 

Throughout, we adopt the framework of a random effects model, although the approaches introduced in this paper are similarly applicable to the common effect model. The classic inverse-variance weighted estimator of $\mu_{\text{pool}}$ and its estimated SE are given by
\begin{equation*}
    \hat{\mu}_{\text{pool}} = \frac{\sum_{k = 1}^K \frac{y_k}{\sigma_k^2 + \hat{\tau}^2}}{\sum_{k = 1}^K \frac{1}{\sigma_k^2 + \hat{\tau}^2}}, \qquad \widehat{\text{SE}}(\hat{\mu}_{\text{pool}}) = \sqrt{\frac{1}{\sum_{k = 1}^K \frac{1}{\sigma_k^2 + \hat{\tau}^2}}}.
\end{equation*}
where $\hat{\tau}^2$ denotes an estimate of the between-study variance. A number of between-study variance estimators have been proposed and systematically compared, which are discussed at length elsewhere \cite{viechtbauer2005bias, veroniki2016methods, langan2019comparison}. A commonly used relative measure of between-study heterogeneity is the $I^2$ index. The $I^2$ index describes the proportion of the variability of the study-specific means that is due to the between-study variance rather than the within-study variance (i.e., $\tau^2 / \Var(y_k)$) \cite{higgins2002quantifying, higgins2003measuring}. The $I^2$ index is typically estimated by
\begin{equation*}
    \hat{I}^2 = \frac{\hat{\tau}^2}{\hat{\tau}^2 + \hat{\sigma}^2}
\end{equation*}
where $\hat{\sigma}^2$ is an estimator of the ``typical" within-study variance. Higgins et al.\ \cite{higgins2002quantifying} discuss approaches for obtaining $\hat{\sigma}^2$.

\subsection{Transformation-based approaches} \label{sec:transformation}

As previously discussed, some primary studies in a meta-analysis may report the sample mean of the outcome of interest and other primary studies may report the median. Transformation-based approaches are conventionally applied in such settings to impute the sample means and standard deviations from primary studies reporting medians. In this subsection, we describe such transformation-based approaches. For simplicity, we consider the setting of one-group primary studies, although the methods described here are applicable to studies with multiple groups (e.g., treatment and control groups) where the outcome of interest may be the difference of means across groups.

We use the following notation to describe data arising in a primary study. Suppose we observe $n$ independent and identically distributed samples of the outcome of interest $X_i$. Let $\mathbb{P}$ denote the underlying distribution and $\mathcal{P}$ denote the underlying model. The primary study may report the following summary statistics of the outcome of interest: minimum value ($q_{\text{min}}$), first quartile ($q_{1}$), median ($q_{2}$), third quartile ($q_{3}$), and maximum value ($q_{\text{max}}$). 

Transformation-based approaches typically consider that one of the following sets of summary statistics is reported:
\begin{align*}
    S_1 & = \{q_{\text{min}}, q_{2}, q_{\text{max}}, n \} \\
    S_2 & = \{q_{1}, q_{2}, q_{3}, n \} \\
    S_3 & = \{q_{\text{min}}, q_{1}, q_{2}, q_{3}, q_{\text{max}}, n \}. 
\end{align*}

In the following subsections, we describe the transformation-based approaches proposed by McGrath et al.\ \cite{mcgrath2020estimating} and Cai et al.\ \cite{cai2021estimating}. We focus on these approaches because they were developed for situations when the data were suspected to be non-normal -- which is often the case when primary studies report the median of the outcome -- and these approaches yielded excellent performance for estimating the mean and standard deviation in simulation studies under non-normal distributions. 

\subsubsection{Quantile Matching Estimation (QE)}
The quantile (matching) estimation (QE) approach \cite{mcgrath2020estimating} assumes that the underlying distribution $\mathbb{P}$ belongs to one of following candidate parametric distributions: normal, log-normal, gamma, beta, or Weibull. This approach estimates the parameters of each candidate distribution by minimizing the distance between the sample quantiles with the distribution quantiles. More specifically, let $\mathcal{P}_{j} = \{ \mathbb{P}_{j, \theta_j} : \theta_j \in \Theta_j \}$ denote the $j$th candidate parametric distribution. This approach obtains an estimate $\hat{\theta}_j$ of $\theta_j$ by minimizing $S_j(\theta_j)$ where
\begin{align*}
    S_j(\theta_j) & := (F^{-1}_{j, \theta_j}(1/n) - q_{\text{min}})^2 + (F^{-1}_{j, \theta_j}(1/2) - q_{2})^2 + (F^{-1}_{j, \theta_j}(1-1/n) - q_{\text{max}})^2 \quad  && \text{in $S_1$}\\
    S_j(\theta_j) & := (F^{-1}_{j, \theta_j}(1/4) - q_{1})^2 + (F^{-1}_{j, \theta_j}(1/2) - q_{2})^2 + (F^{-1}_{j, \theta_j}(3/4) - q_{3})^2  && \text{in $S_2$}\\
    S_j(\theta_j) & := (F^{-1}_{j, \theta_j}(1/n) - q_{\text{min}})^2 + (F^{-1}_{j, \theta_j}(1/4) - q_{1})^2 + (F^{-1}_{j, \theta_j}(1/2) - q_{2})^2 \\ & \qquad \qquad \qquad + (F^{-1}_{j, \theta_j}(3/4) - q_{3})^2 + (F^{-1}_{j, \theta_j}(1-1/n) - q_{\text{max}})^2 && \text{in $S_3$}
\end{align*}
and $F^{-1}_{j, \theta_j}$ denotes the inverse cumulative distribution function (CDF) of distribution $\mathbb{P}_{j,\theta_j}$.
The candidate distribution with the best fit (i.e., smallest value of $S_j(\hat{\theta}_j)$) is selected. The mean and standard deviation of the selected distribution are used as the estimated mean and standard deviation.

\subsubsection{Box-Cox (BC)}
The Box-Cox (BC) approach \cite{mcgrath2020estimating} assumes that the underlying distribution is normal after applying a suitable Box-Cox transformation \cite{box1964analysis}. A Box-Cox transformation with power parameter $\lambda$ is defined as 
\begin{equation*}
    g_{\lambda}(x) := \begin{cases} \frac{x^\lambda - 1}{\lambda} \qquad & \text{if $\lambda \neq 0$} \\ \log x & \text{if $\lambda = 0$} \end{cases}
\end{equation*}
for $x > 0$. In $S_1$, the BC approach estimates $\lambda$ by finding the value such that the transformed minimum and maximum values (i.e., $g_\lambda(q_{\text{min}})$ and $g_\lambda(q_{\text{max}})$) are equidistant from the transformed median (i.e., $g_\lambda(q_2)$). Similarly, in $S_2$, $\lambda$ is estimated such that the transformed first and third quartiles are equidistant from the transformed median. Since there does not necessarily exist a value of $\lambda$ such that both (i) the transformed minimum and maximum values are equidistant from the transformed median, and (ii) the transformed first and third quartiles are equidistant from the transformed median, the BC approach estimates $\lambda$ in $S_3$ by 
\begin{equation*}
    \hat{\lambda} = \argmin_\lambda \left\{ \left[ (g_{\lambda}(q_3) - g_{\lambda}(q_2)) - (g_{\lambda}(q_2) - g_{\lambda}(q_1)) \right]^2 + \left[ (g_{\lambda}(q_3) - g_{\lambda}(q_2)) - (g_{\lambda}(q_2) - g_{\lambda}(q_1)) \right]^2 \right\}.
\end{equation*}
Then, the methods of Luo et al.\ \cite{luo2018optimally} and Wan et al.\ \cite{wan2014estimating} are applied to estimate the mean and standard deviation of the transformed data, respectively. Last, the inverse transformation is applied to estimate the mean and standard deviation of the original, untransformed data.

\subsubsection{Method for Unknown Non-Normal Distributions (MLN)}

Like the BC approach, the Method for Unknown Non-Normal Distributions (MLN) \cite{cai2021estimating} assumes that the underlying distribution is normal after applying a suitable Box-Cox transformation (i.e., $g_\lambda(X_i) \overset{\text{i.i.d.}}{\sim} \text{Normal}(\mu, \sigma^2)$). The following maximum likelihood approach is used to estimate $\lambda$. For $s = 1, 2, 3$, let $(\tilde{\mu}_s, \tilde{\sigma}_s)$ denote the estimated mean and standard deviation based on the methods of Luo et al.\ \cite{luo2018optimally} and Wan et al.\ \cite{wan2014estimating} applied to the Box-Cox transformed quantiles of summary statistics $S_s$ with power parameter $\lambda$ (which we omit for notational simplicity). Let $F_s$ denote the CDF and $f_s$ denote the probability density function of a $\text{Normal}(\tilde{\mu}_s, \tilde{\sigma}_s^2)$ distribution, $s = 1, 2, 3$. The conditional likelihood in $S_1$ is given by
\begin{equation*}
     L(\lambda | S_1, \tilde{\mu}_1, \tilde{\sigma}_1) \propto (F_1(q_{2}) - F_1(q_{\text{min}}))^{n/2-1}(F_1(q_{\text{max}}) - F_1(q_{2}))^{n/2-1} f_1(q_{\text{min}}) f_1(q_{2}) f_1(q_{\text{max}}).
\end{equation*}
Similarly, the conditional likelihood in $S_2$ is given by
\begin{equation*} 
    L(\lambda | S_2, \tilde{\mu}_2, \tilde{\sigma}_2) \propto F_2(q_{1})^{n/4}(F_2(q_{2}) - F_2(q_{1}))^{n/4-1}(F_2(q_{3}) - F_2(q_{2}))^{n/4-1}(1 - F_2(q_{3}))^{n/4} f_2(q_{1}) f_2(q_{2}) f_2(q_{3})
\end{equation*}
and in $S_3$ is given by
\begin{align*}
    L(\lambda | S_3, \tilde{\mu}_3, \tilde{\sigma}_3) & \propto (F_3(q_{1}) - F_3(q_{\text{min}}))^{n/4-1}(F_3(q_{2}) - F_3(q_{1}))^{n/4-1}(F_3(q_{3}) - F_3(q_{2}))^{n/4-1}(F_3(q_{\text{max}}) - F_3(q_{3}))^{n/4-1} \\
    & \qquad \qquad \qquad \times f_3(q_{\text{min}}) f_3(q_{1}) f_3(q_{2}) f_3(q_{3}) f_3(q_{\text{max}}).
\end{align*}
This approach estimates $\lambda$ by minimizing the log conditional likelihood. Then, this approach finds the maximum likelihood estimate of $\mu$ and $\sigma$ conditional on the sample quantiles and $\hat{\lambda}$. Last, the inverse transformation is applied to estimate the mean and standard deviation of the untransformed data.

\subsection{Standard error estimation of transformation-based approaches}

As a slight abuse of notation, in this subsection we let $\sigma$ denote the standard deviation of the underlying distribution of the outcome in a given study. Let $\hat{\sigma}_{\text{sample}}$ denote the sample standard deviation of the outcome and $\hat{\sigma}_{\text{trans}}$ denote the estimated standard deviation of the outcome based on an arbitrary transformation-based approach.

\subsubsection{Naïve approach} \label{sec:naive SE}

Recall from Section \ref{sec:models} that standard meta-analytic approaches require that each primary study contributes a point estimate of the outcome of interest and an estimate of its SE. When a primary study reports the sample mean of the outcome of interest, the usual SE estimate is $\widehat{\text{SE}} = \hat{\sigma}_{\text{sample}} / \sqrt{n}$ (as the true SE is $\sigma / \sqrt{n}$). When primary studies report medians, data analysts typically apply transformation-based approaches and use the estimated mean as the point estimate and $\hat{\sigma}_{\text{trans}} / \sqrt{n}$ as its SE estimate. 

However, $\sigma / \sqrt{n}$ is not necessarily equal to the true SE of the mean estimator of the transformation-based approaches. That is, $\sigma / \sqrt{n}$ is the wrong target parameter for the within-study SE when using transformation-based approaches. Unlike the sample mean, the mean estimator of the transformation-based approaches is not based on the full data set. Consequently, one may expect the true SE of the mean estimator of transformation-based approaches to be systematically higher than $\sigma / \sqrt{n}$. Further, one may expect this discrepancy to be particularly large for the more complicated transformation-based approaches involving model selection and/or solving various optimization problems, such as those described in the previous subsections. For these reasons, we refer to $\hat{\sigma}_{\text{trans}} / \sqrt{n}$ as the naïve SE estimate of the transformation-based approaches.

\begin{figure} [H]
    \centering
    \includegraphics[width=\textwidth]{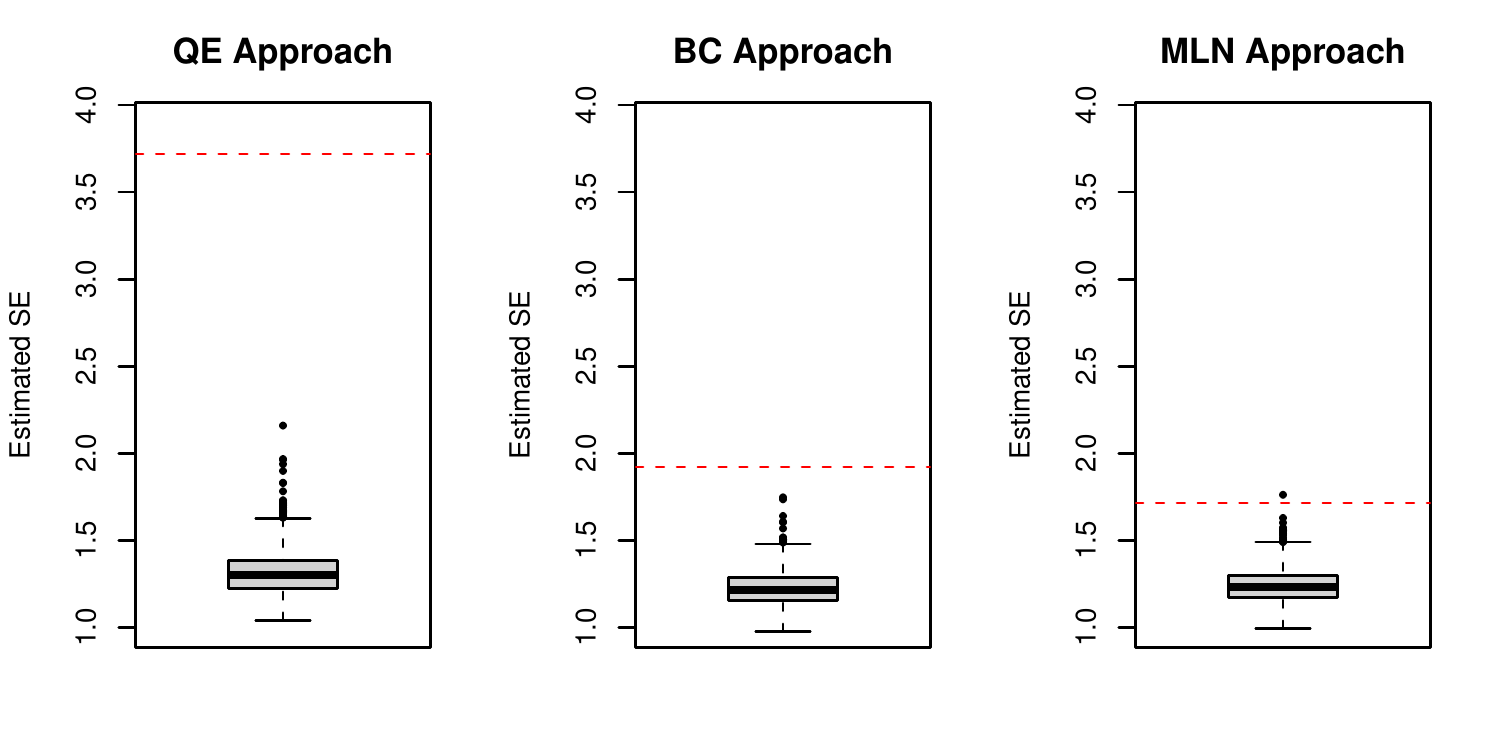}
    \caption{Boxplot of the naïve SE estimates of the QE, BC, and MLN approaches. The dotted red lines illustrate the true SEs of the three approaches. \label{fig:illustration}}
\end{figure}

As a simple illustration of the bias of the naïve SE estimates, consider the following example. We generated 1,000 independent data sets of size $n = 1,000$ under a $\text{Log-Normal}(5, 0.25^2)$ distribution. For each data set, we calculated $S_1$ summary statistics and applied the QE, BC, and MLN approaches to the summary data. Figure \ref{fig:illustration} displays the distributions of the naïve SE estimates along with the true SEs of the three approaches (obtained by Monte Carlo integration with $10^5$ samples). The naïve SE estimators of all three approaches severely underestimated the true SEs. This bias is not due to poor estimation of $\sigma$: The value of $\sigma / \sqrt{n}$ (i.e., representing the best-case scenario for the naïve approaches) is approximately 1.2, which is considerably smaller than the true SEs. Similar results hold when applying other transformation-based approaches to this example (see Appendix \ref{sec: appendix illustrative example}).

\subsubsection{Parametric bootstrap}

To remedy the problems of the naïve approaches, we propose a straightforward application of parametric bootstrap \cite{efron1994introduction} to estimate the SEs of the transformation-based approaches. For the sake of concreteness, we describe the parametric bootstrap approach for the MLN method under $S_1$, noting that only trivial modifications are needed for the other transformation-based methods and scenarios (i.e., $S_2$, $S_3$).

First, estimate the parameters of the underlying distribution by applying the MLN approach to the $S_1$ summary data. Let $\hat{\mathbb{P}}$ denote the fitted distribution. For $b = 1, \dots, B$ where $B$ is some large number (e.g., $B = 1000)$, perform the following two steps: (1) Simulate $n$ i.i.d.\ observations under $\hat{\mathbb{P}}$, (2) Apply the MLN approach to obtain an estimate of the mean ($\hat{\theta}_b$) based on minimum value, median, and maximum value of the simulated data in the first step. The parametric bootstrap estimate of the SE of the MLN approach is the sample standard deviation of the $\hat{\theta}_b$ (i.e., $\sqrt{\frac{1}{B-1} \sum_{b = 1}^B (\hat{\theta}_b - \overline{\theta})^2}$ where $\overline{\theta} = \frac{1}{B} \sum_{b = 1}^B \hat{\theta}_b$). 

The distributions of the SE estimates when applying the parametric bootstrap approach to the illustrative example in the previous subsection are given in Appendix \ref{sec: appendix illustrative example}.

\section{Simulations} \label{sec:simulations}
We systematically evaluated the performance of the naïve and parametric bootstrap SE estimators of the QE, BC, and MLN approaches in a simulation study. We performed simulations at both the study-level (Section \ref{sec:study level}) and at the meta-analytic level (Section \ref{sec:meta-analytic level}). The code used for these simulations is available at \url{https://github.com/stmcg/se-est}.

\subsection{Study-level simulations} \label{sec:study level}

\subsubsection{Simulation Settings}

For consistency with previous simulation studies evaluating the performance of the QE, BC, and MLN methods \cite{mcgrath2020estimating, cai2021estimating}, we generated independent and identically distributed data sets of size $n$ under the following distributions: $\text{Log-Normal}(5, 0.25^2)$, $\text{Log-Normal}(5, 0.5^2)$, $\text{Log-Normal}(5, 1)$, and $\text{Normal}(5, 1)$. Unlike previous studies, we also included the half-normal distribution with a mean of 10 to evaluate the performance of the all three approaches under model misspecification. The probability density functions of these five distributions are illustrated in Appendix \ref{sec: appendix study level}. We also varied the sample size ($n \in \{50, 250, 1000\}$) and the summary statistics reported (i.e., $S_1$, $S_2$, or $S_3$). We performed $N=1000$ repetitions for each of the $5 \times 3 \times 3 = 45$ combinations of distribution, sample size, and summary statistics reported.  

We applied the QE, BC, and MLN approaches to obtain the naïve and parametric bootstrap SE estimates. We used $B=1000$ bootstrap samples.

We evaluated the median percent error, mean percent error, and root mean squared error (RMSE) of the SE estimators. Letting $\widehat{\text{SE}}_i$ denote an estimate of the SE in the $i$th repetition ($i = 1, \dots, N$), the median percent error is defined to be the sample median of the $(\widehat{\text{SE}}_i - \text{SE}) / \text{SE} \times 100\%$ and the mean percent error is defined to be the sample mean of the $(\widehat{\text{SE}}_i - \text{SE}) / \text{SE}  \times 100\%$. The true values of the SEs were obtained by Monte Carlo integration with $10^5$ samples.

\subsubsection{Results}

The median percent error of the naïve and bootstrap SE estimators in scenarios $S_1$, $S_2$, and $S_3$ are given in Tables \ref{table:studylevel mre S1}, \ref{table:studylevel mre S2}, and \ref{table:studylevel mre S3}, respectively. The naïve approaches underestimated the SEs in all simulation settings. These approaches generally performed worse as $n$ increased, which may be expected since these approaches estimate the wrong target parameter. In one of the least favorable simulation settings for the naïve approaches, the naïve SE estimators of the QE, BC, and MLN approaches had median percent error of -70\%, -59\%, and -54\%, respectively, under the $\text{Log-Normal}(5, 1)$ in scenario $S_1$ when $n = 1000$. 

The bootstrap SE estimators had considerably smaller median percent error compared to the naïve approaches in most simulation settings. These approaches also generally performed better as $n$ increased. There was not a clear trend in the performance of these methods when varying the summary statistics reported (i.e., $S_1$, $S_2$, $S_3$). In general, the bootstrap SE estimator of the MLN approach performed best. For instance, this approach had median percent error smaller than 2\% in most simulation settings.

The simulation results for the mean percent error and RMSE are given in Appendix \ref{sec: appendix study level} for parsimony. We observed similar trends: the bootstrap approaches often performed better than the naïve approaches. However, there were occasionally very large bootstrap SE estimates which resulted in the bootstrap approaches having larger mean percent error and RMSE compared to the naïve approach in some simulation settings. This occurred most often for the QE approach (especially in $S_3$ with small $n$).

\begin{table}[H]
\caption{Median percent error of the naïve and parametric bootstrap SE estimators in scenario $S_1$. The median percent error values are rounded to the nearest integer. \label{table:studylevel mre S1}}
\begin{center}
\begin{tabular}{@{\extracolsep{6pt}}llllllll@{}}
\hline
& & \multicolumn{2}{c}{QE Approach} & \multicolumn{2}{c}{BC Approach} & \multicolumn{2}{c}{MLN Approach} \\ \cline{3-4} \cline{5-6} \cline{7-8} 
 Distribution & $n$ & Naïve & Boot.\ & Naïve & Boot. & Naïve  & Boot.\ \\
  \hline
$\text{Log-Normal}(5, 0.25^2)$  & 50 &  -22 & 6 & -19 & -4 & -16 & -2 \\ 
& 250 & -47 & 3 & -27 & -2 & -22 & 0 \\ 
& 1000 & -65 & 4 & -37 & -1 & -28 & 1 \\ 
$\text{Log-Normal}(5, 0.5^2)$ & 50 & -23 & 10 & -24 & -6 & -19 & -4 \\ 
& 250 & -42 & 8 & -33 & -4 & -28 & 0 \\ 
& 1000 & -59 & 5 & -46 & -3 & -40 & 0 \\ 
$\text{Log-Normal}(5, 1)$ & 50 & -45 & 37 & -33 & -12 & -24 & -4 \\ 
& 250 & -57 & 18 & -46 & -8 & -39 & -2 \\ 
& 1000 & -70 & 13 & -59 & -3 & -54 & 3 \\ 
$\text{Normal}(5, 1)$ & 50 & -19 & 12 & -17 & -2 & -11 & 4 \\
& 250 & -44 & 12 & -25 & -1 & -20 & 1 \\ 
& 1000 & -71 & 2 & -35 & -1 & -29 & 2 \\ 
$\text{Half-Normal}(0.1)$ & 50 & -18 & 9 & -27 & -7 & -18 & 5 \\  
& 250 & -32 & 6 & -37 & -7 & -32 & 1 \\ 
& 1000 & -46 & 10 & -47 & -2 & -42 & 5 \\ 
   \hline
\end{tabular}
\end{center}
\end{table}

\begin{table}[H]
\caption{Median percent error of the naïve and parametric bootstrap SE estimators in scenario $S_2$. The median percent error values are rounded to the nearest integer.\label{table:studylevel mre S2}}
\begin{center}
\begin{tabular}{@{\extracolsep{6pt}}llllllll@{}}
\hline
& & \multicolumn{2}{c}{QE Approach} & \multicolumn{2}{c}{BC Approach} & \multicolumn{2}{c}{MLN Approach} \\ \cline{3-4} \cline{5-6} \cline{7-8} 
 Distribution & $n$ & Naïve & Boot.\ & Naïve & Boot. & Naïve  & Boot.\ \\
  \hline
$\text{Log-Normal}(5, 0.25^2)$  & 50 &  -13 & -4 & -20 & -1 & -13 & -6 \\ 
& 250 & -19 & -2 & -29 & 1 & -10 & -1 \\ 
& 1000 & -27 & 1 & -35 & 3 & -13 & -1 \\ 
$\text{Log-Normal}(5, 0.5^2)$ & 50 & -21 & -8 & -28 & -8 & -18 & -9 \\ 
& 250 & -23 & -5 & -34 & -2 & -15 & -2 \\ 
& 1000 & -29 & -1 & -38 & 2 & -16 & 0 \\ 
$\text{Log-Normal}(5, 1)$ & 50 & -42 & -19 & -46 & -18 & -27 & -15 \\ 
& 250 & -30 & -2 & -43 & -5 & -21 & -4 \\ 
& 1000 & -22 & -1 & -46 & -1 & -22 & -1 \\ 
$\text{Normal}(5, 1)$ & 50 & -12 & -2 & -19 & 2 & -10 & -2 \\ 
& 250 & -23 & -3 & -34 & -3 & -6 & 3 \\ 
& 1000 & -38 & -6 & -46 & -3 & -11 & 1 \\ 
$\text{Half-Normal}(0.1)$ & 50 & -36 & -6 & -50 & -3 & 0 & 16 \\ 
& 250 & -49 & 4 & -56 & -1 & -28 & 2 \\ 
& 1000 & -52 & -5 & -60 & 1 & -57 & 8 \\ 
   \hline
\end{tabular}
\end{center}
\end{table}

\begin{table}[H]
\caption{Median percent error of the naïve and parametric bootstrap SE estimators in scenario $S_3$. The median percent error values are rounded to the nearest integer. \label{table:studylevel mre S3}}
\begin{center}
\begin{tabular}{@{\extracolsep{6pt}}llllllll@{}}
\hline
& & \multicolumn{2}{c}{QE Approach} & \multicolumn{2}{c}{BC Approach} & \multicolumn{2}{c}{MLN Approach} \\ \cline{3-4} \cline{5-6} \cline{7-8} 
 Distribution & $n$ & Naïve & Boot.\ & Naïve & Boot. & Naïve  & Boot.\ \\
  \hline
$\text{Log-Normal}(5, 0.25^2)$  & 50 &  -5 & 9 & -8 & -1 & -7 & -3 \\ 
& 250 & -24 & 4 & -15 & -1 & -8 & 0 \\ 
& 1000 & -43 & 3 & -26 & 0 & -12 & 0 \\ 
$\text{Log-Normal}(5, 0.5^2)$ & 50 & -8 & 11 & -13 & -4 & -9 & -5 \\ 
& 250 & -20 & 7 & -19 & -2 & -10 & -1 \\ 
& 1000 & -36 & 7 & -30 & -1 & -15 & -1 \\ 
$\text{Log-Normal}(5, 1)$ & 50 & -43 & 44 & -19 & -6 & -10 & -7 \\ 
& 250 & -44 & 16 & -25 & -5 & -13 & -3 \\ 
& 1000 & -44 & 11 & -39 & -3 & -18 & -1 \\ 
$\text{Normal}(5, 1)$ & 50 & -3 & 8 & -7 & 0 & -2 & 2 \\ 
& 250 & -21 & 7 & -15 & 0 & -8 & -1 \\ 
& 1000 & -51 & 1 & -25 & 0 & -16 & 0 \\ 
$\text{Half-Normal}(0.1)$ & 50 & -5 & 7 & -15 & 9 & 4 & 18 \\ 
& 250 & -16 & -2 & -20 & 3 & -7 & 11 \\ 
& 1000 & -22 & 1 & -30 & 4 & -13 & 19 \\ 
   \hline
\end{tabular}
\end{center}
\end{table}

\subsection{Meta-analysis simulations} \label{sec:meta-analytic level}

\subsubsection{Simulation settings}

We simulated meta-analyses with primary studies reporting sample medians of the outcome of interest as follows. We simulated the outcome of interest for the $i$th individual in the $k$th primary study $(X_{ik})$ by
\begin{align*}
    X_{ik} & \sim \text{Log-Normal}(5, 0.25^2) + \gamma_k, \qquad i = 1, \dots, n_k, \, k = 1, \dots, K 
\end{align*}
where $\gamma_k$ denotes a study-specific random effect. We let $\gamma_k \sim \text{Normal}(0, \tau^2)$ for $k = 1, \dots, K$ where $\tau^2$ was set to 6. This value of $\tau^2$ was set to obtain a reasonable degree of between-study heterogeneity (e.g., $I^2 \approx 50\%$ if all studies report sample means). The study-specific sample sizes were drawn from a discrete uniform distribution with a minimum value of $100$ and a maximum value of $500$. 

We included 20 different simulation settings by varying number of primary studies $(k \in \{9, 30\})$ the proportion of primary studies reporting the sample median of the outcome ($p \in \{0, \frac{1}{3}, \frac{2}{3}, 1\}$) and the summary statistics reported by the primary studies reporting medians (i.e., $S_1, S_2, S_3$). When primary studies did not report the sample median of the outcome, they reported the sample mean, standard deviation, and sample size. We performed $1000$ repetitions in each of the 20 simulation settings.

We applied the QE, BC, and MLN approaches to all primary studies reporting medians. We applied each of the three transformation-based approaches with the naïve SE estimator and the parametric bootstrap SE estimator with $B = 1000$ bootstrap samples. After obtaining all study-specific points estimates and estimates of their SEs, we applied the standard inverse-variance approach to estimate the pooled mean. We used the Restricted Maximum Likelihood (REML) approach to estimate the between-study variance. To construct 95\% CIs around the pooled mean and between-study variance estimates, we used the Wald method for the pooled mean and the Q-profile method for the between-study variance, as implemented in the \texttt{metafor} R package \cite{metafor}. 

We evaluated the bias, variance, and coverage of the 95\% confidence intervals (CIs) of the pooled mean and between-study variance estimators. We also evaluated the bias of the $I^2$ estimators. The true values of $I^2$ were obtained by Monte Carlo integration with $10^6$ samples.

Note that we included two simulation settings where all primary studies report the sample mean of the outcome (i.e., settings with $p = 0$), in which case transformation-based approaches are not applicable. We included these settings to clarify the extent to which sub-optimal performance of the estimators (e.g., any bias or below nominal coverage of 95\% CIs) may be attributed to sources such as a small number of primary studies or small within-study sample sizes rather than poor performance of the transformation-based approaches.

\subsubsection{Results}

\paragraph{Pooled mean estimators} 

The bias, variance, and coverage of the 95\% CIs of the pooled mean estimators are given in Appendix \ref{sec: appendix meta-analytic level}. The pooled mean estimators using the transformation-based approaches with bootstrap SE estimates often had similar bias, variance, and coverage compared to when using the naïve SE estimates. The main exception to these trends was the QE approach in $S_1$, which had smaller bias and variance as well as better coverage when using bootstrap SE estimates compared to the naïve SE estimates.

\paragraph{Between-study variance estimators} Tables \ref{table:meta-analysis tausq bias} and \ref{table:meta-analysis tausq cov} give the bias and coverage of the 95\% CIs of the between-study variance estimators. The variance of the between-study variance estimators is given in Appendix \ref{sec: appendix meta-analytic level}.

The naïve approaches overestimated the between-study variance in all simulation scenarios, especially when the proportion of studies reporting medians was large. Consequently, they did not usually attain nominal coverage of their 95\% CIs for between-study variance. As one may expect, the coverage of the naïve approaches decreased as the total number of primary studies and the proportion of primary studies reporting medians increased.  

The bootstrap approaches performed considerably better than the naïve approaches for estimating the between-study variance. These approaches had smaller bias and variance, and they usually attained nominal or near-nominal coverage of their 95\% CIs. There were not clear trends in the performance of these approaches when varying the summary statistics reported (i.e., $S_1$, $S_2$, $S_3$), the number of primary studies, or the proportion of primary studies reporting medians. In general, the MLN approach with bootstrap SEs performed best for estimating the between-study variance.

\begin{table}[H]
\caption{Bias of the $\tau^2$ estimators. The number of primary studies is denoted by $K$ and the proportion of primary studies reporting medians is denoted by $p$. The last two rows correspond to the scenario where all studies report the sample mean and standard deviation of the outcome, in which case the transformation-based approaches are not applicable. The true value of $\tau^2$ is 6. \label{table:meta-analysis tausq bias}}
\begin{center}
\begin{tabular}{@{\extracolsep{6pt}}lllllllll@{}}
\hline
& & & \multicolumn{2}{c}{QE Approach} & \multicolumn{2}{c}{BC Approach} & \multicolumn{2}{c}{MLN Approach} \\ \cline{4-5} \cline{6-7} \cline{8-9} 
Scenario & $K$ & $p$ & Naïve & Boot. & Naïve & Boot. &  Naïve & Boot. \\
  \hline
$S_1$ & 9 & $1/3$   &  5.96 & 0.35 & 1.81 & 0.44 & 1.29 & 0.28 \\ 
& & $2/3$   & 11.67 & 0.85 & 3.34 & 0.69 & 2.28 & 0.32 \\ 
& & $1$   & 15.90 & 0.61 & 4.88 & 0.96 & 3.50 & 0.54 \\ 
 & 30 & $1/3$   &  5.40 & -0.05 & 1.57 & 0.14 & 1.06 & 0.02 \\ 
& & $2/3$   & 10.79 & -0.33 & 3.00 & 0.17 & 2.09 & -0.03 \\ 
& & $1$   & 15.15 & -1.83 & 4.09 & -0.09 & 2.77 & -0.45 \\ 
$S_2$ & 9 & $1/3$   &  1.24 & 0.45 & 2.20 & 0.44 & 0.43 & 0.08 \\ 
& & $2/3$   & 1.93 & 0.45 & 3.65 & 0.33 & 0.63 & -0.05 \\ 
& & $1$   & 2.40 & 0.20 & 4.78 & 0.12 & 1.05 & 0.06 \\ 
 & 30 & $1/3$   &  1.21 & 0.34 & 2.17 & 0.25 & 0.48 & 0.09 \\ 
& & $2/3$   & 1.84 & 0.23 & 3.66 & 0.15 & 0.62 & -0.11 \\ 
& & $1$   & 2.61 & 0.18 & 5.25 & -0.09 & 1.16 & 0.04 \\ 
$S_3$ & 9 & $1/3$   &  1.45 & -0.08 & 0.74 & 0.14 & 0.30 & 0.03 \\ 
& & $2/3$   & 3.77 & 0.64 & 1.79 & 0.57 & 1.10 & 0.52 \\ 
& & $1$   & 4.19 & -0.37 & 2.06 & 0.25 & 0.89 & 0.07 \\ 
& 30 & $1/3$   &  1.68 & 0.02 & 0.90 & 0.20 & 0.44 & 0.13 \\ 
& & $2/3$   & 2.87 & -0.30 & 1.58 & 0.24 & 0.71 & 0.12 \\ 
& & $1$   & 4.23 & -0.98 & 2.11 & 0.09 & 0.85 & -0.06 \\  \hdashline 
& 9 & $0$ & 0.08 & 0.08 & 0.08 & 0.08 & 0.08 & 0.08 \\ 
& 30 & $0$ & -0.05 & -0.05 & -0.05 & -0.05 & -0.05 & -0.05 \\ 
   \hline
\end{tabular}
\end{center}
\end{table}

\begin{table}[H]
\caption{Coverage of the 95\% confidence intervals for $\tau^2$. The number of primary studies is denoted by $K$ and the proportion of primary studies reporting medians is denoted by $p$. The last two rows correspond to the scenario where all studies report the sample mean and standard deviation of the outcome, in which case the transformation-based approaches are not applicable. \label{table:meta-analysis tausq cov}}
\begin{center}
\begin{tabular}{@{\extracolsep{6pt}}lllllllll@{}}
\hline
& & & \multicolumn{2}{c}{QE Approach} & \multicolumn{2}{c}{BC Approach} & \multicolumn{2}{c}{MLN Approach} \\ \cline{4-5} \cline{6-7} \cline{8-9} 
Scenario & $K$ & $p$ & Naïve & Boot. & Naïve & Boot. &  Naïve & Boot. \\
  \hline
$S_1$ & 9 & $1/3$   & 0.82 & 0.94 & 0.92 & 0.94 & 0.93 & 0.95 \\ 
& & $2/3$   & 0.66 & 0.95 & 0.89 & 0.95 & 0.92 & 0.94 \\ 
& & $1$   & 0.55 & 0.96 & 0.86 & 0.94 & 0.90 & 0.95 \\ 
& 30 & $1/3$   &  0.61 & 0.94 & 0.90 & 0.95 & 0.92 & 0.94 \\ 
& & $2/3$   & 0.30 & 0.94 & 0.82 & 0.93 & 0.87 & 0.93 \\ 
& & $1$   & 0.13 & 0.94 & 0.74 & 0.95 & 0.84 & 0.96 \\ 
$S_2$ & 9 & $1/3$   &  0.94 & 0.95 & 0.91 & 0.94 & 0.95 & 0.95 \\ 
& & $2/3$   & 0.93 & 0.96 & 0.90 & 0.95 & 0.94 & 0.96 \\ 
& & $1$   & 0.93 & 0.96 & 0.87 & 0.95 & 0.94 & 0.95 \\ 
& 30 & $1/3$   & 0.91 & 0.94 & 0.87 & 0.93 & 0.94 & 0.94 \\ 
& & $2/3$   & 0.89 & 0.95 & 0.77 & 0.94 & 0.95 & 0.96 \\ 
& & $1$   & 0.84 & 0.96 & 0.63 & 0.95 & 0.92 & 0.94 \\ 
$S_3$ & 9 & $1/3$   &  0.94 & 0.96 & 0.94 & 0.94 & 0.95 & 0.95 \\ 
& & $2/3$   & 0.87 & 0.95 & 0.94 & 0.95 & 0.94 & 0.95 \\ 
& & $1$   & 0.90 & 0.96 & 0.92 & 0.95 & 0.94 & 0.95 \\ 
& 30 & $1/3$   &  0.90 & 0.96 & 0.94 & 0.96 & 0.95 & 0.96 \\ 
& & $2/3$   & 0.84 & 0.95 & 0.92 & 0.96 & 0.95 & 0.95 \\ 
& & $1$   & 0.77 & 0.95 & 0.88 & 0.94 & 0.92 & 0.94 \\ 
\hdashline 
& 9 & $0$ & 0.95 & 0.95 & 0.95 & 0.95 & 0.95 & 0.95 \\ 
& 30 & $0$ & 0.95 & 0.95 & 0.95 & 0.95 & 0.95 & 0.95 \\ 
   \hline
\end{tabular}
\end{center}
\end{table}

\paragraph{$I^2$ estimators}

The bias of the $I^2$ estimators are given in Table \ref{table:meta-analysis I2 bias}. We observed similar trends as those for estimating the between-study variance. 

More specifically, the naïve approaches overestimated $I^2$ in all scenarios, especially when the proportion of studies reporting medians was large and $S_1$ summary statistics were reported. For instance, the QE, BC, and MLN naïve approaches overestimated $I^2$ by 58, 30, and 22 percentage points, respectively, when the number of primary studies was 30 and all primary studies reported $S_1$ summary statistics. The bootstrap approaches often performed considerably better. These approaches had bias smaller than 5 percentage points in most scenarios.

There were some scenarios (e.g., often when $K = 9$) in which the MLN approach with naïve SEs had smaller bias for estimating $I^2$ compared to when using the bootstrap SEs. This typically occurred when the bias of both the naïve the bootstrap MLN approaches was small. When evaluating the median bias of the $I^2$ estimators (i.e., median of $\hat{I}^2 - I^2$), this phenomenon rarely occurred. See Appendix \ref{sec: appendix meta-analytic level} for details.

\begin{table}[H]
\caption{Bias of the $I^2$ estimators. The number of primary studies is denoted by $K$ and the proportion of primary studies reporting medians is denoted by $p$. The last two rows correspond to the scenario where all studies report the sample mean and standard deviation of the outcome, in which case the transformation-based approaches are not applicable. The true values of $I^2$ depend on the summary statistics reported, $p$, and the transformation-based approach. The true values of $I^2$ ranged from 19\% to 50\%. \label{table:meta-analysis I2 bias}}
\begin{center}
\begin{tabular}{@{\extracolsep{6pt}}lllllllll@{}}
\hline
& & & \multicolumn{2}{c}{QE Approach} & \multicolumn{2}{c}{BC Approach} & \multicolumn{2}{c}{MLN Approach} \\ \cline{4-5} \cline{6-7} \cline{8-9} 
Scenario & $K$ & $p$ & Naïve & Boot. & Naïve & Boot. &  Naïve & Boot. \\
  \hline
$S_1$ & 9 & $1/3$   & 26.93 & 4.69 & 6.15 & -3.53 & 2.58 & -4.92 \\ 
& & $2/3$   & 43.22 & 3.23 & 15.52 & -2.94 & 9.23 & -5.53 \\ 
& & $1$   & 52.28 & -3.93 & 23.95 & -2.88 & 16.39 & -5.34 \\ 
 & 30 & $1/3$   &  33.98 & 11.37 & 13.33 & 3.24 & 9.79 & 1.89 \\ 
& & $2/3$   & 49.64 & 6.89 & 22.49 & 2.53 & 16.70 & 0.57 \\ 
& & $1$   & 57.50 & -6.83 & 29.57 & -0.31 & 22.35 & -2.50 \\ 
$S_2$ & 9 & $1/3$   &  3.07 & -3.13 & 9.20 & -2.31 & -2.46 & -5.58 \\ 
& & $2/3$   & 8.43 & -3.60 & 18.63 & -3.57 & -0.98 & -6.89 \\ 
& & $1$   & 12.60 & -5.69 & 24.82 & -6.66 & 2.58 & -6.32 \\ 
 & 30 & $1/3$   &  10.20 & 3.36 & 16.48 & 4.16 & 4.77 & 1.39 \\ 
& & $2/3$   & 15.72 & 2.66 & 25.78 & 2.58 & 7.14 & 0.58 \\ 
& & $1$   & 20.55 & 0.83 & 33.20 & -2.14 & 10.53 & 0.65 \\ 
$S_3$ & 9 & $1/3$   &  6.08 & -4.36 & -0.80 & -5.91 & -4.11 & -6.57 \\ 
& & $2/3$   & 16.11 & -3.09 & 5.37 & -4.93 & 0.06 & -5.01 \\ 
& & $1$   & 20.43 & -8.69 & 8.79 & -6.44 & 1.17 & -6.30 \\ 
& 30 & $1/3$   &  14.54 & 3.74 & 7.87 & 2.19 & 4.44 & 1.71 \\ 
& & $2/3$   & 22.11 & 0.77 & 12.71 & 1.54 & 6.60 & 1.23 \\ 
& & $1$   & 28.36 & -6.36 & 17.00 & 0.12 & 8.31 & 0.06 \\  \hdashline 
& 9 & $0$ & -5.60 & -5.60 & -5.60 & -5.60 & -5.60 & -5.60 \\ 
& 30 & $0$ & 1.12 & 1.12 & 1.12 & 1.12 & 1.12 & 1.12 \\ 
   \hline
\end{tabular}
\end{center}
\end{table}

\section{Data application}\label{sec:application}

We performed an empirical comparison of the naïve and parametric bootstrap SE estimators in a meta-analysis of clinical indicators of a severe course of COVID-19 \cite{katzenschlager2021can}. One of the primary analyses of Katzenschlager et al.\ \cite{katzenschlager2021can} compared demographic variables, clinical values, and laboratory values between COVID-19 infected patients who died and those who survived. The original analyses estimated the pooled difference of \emph{medians} of 26 continuous outcomes across the two groups of patients and reported estimates of $I^2$ to describe between-study heterogeneity. 

In our analyses, we applied transformation-based approaches to estimate the pooled difference of \emph{means} of continuous outcomes between COVID-19 infected patients who died and those who survived. All primary studies reported $S_2$ summary statistics or reported the sample mean, standard deviation, and sample size for continuous outcomes. A detailed description of our data processing is given in Appendix \ref{sec: appendix application}. 

We meta-analyzed all outcomes that had at least 6 primary studies reporting appropriate summary data after performing the data processing. We applied the QE, BC, and MLN approaches with the naïve and bootstrap approaches for estimating the within-study SEs. As used in the simulation study, we used $B=1000$ bootstrap samples and used REML to estimate between-study variance. For each outcome, we estimated the pooled difference of means and its 95\% CI, between-study variance, and $I^2$. The data and code used for these analyses are available at \url{https://github.com/stmcg/se-est}.

\begin{figure} [H]
    \centering
    \includegraphics[width=\textwidth]{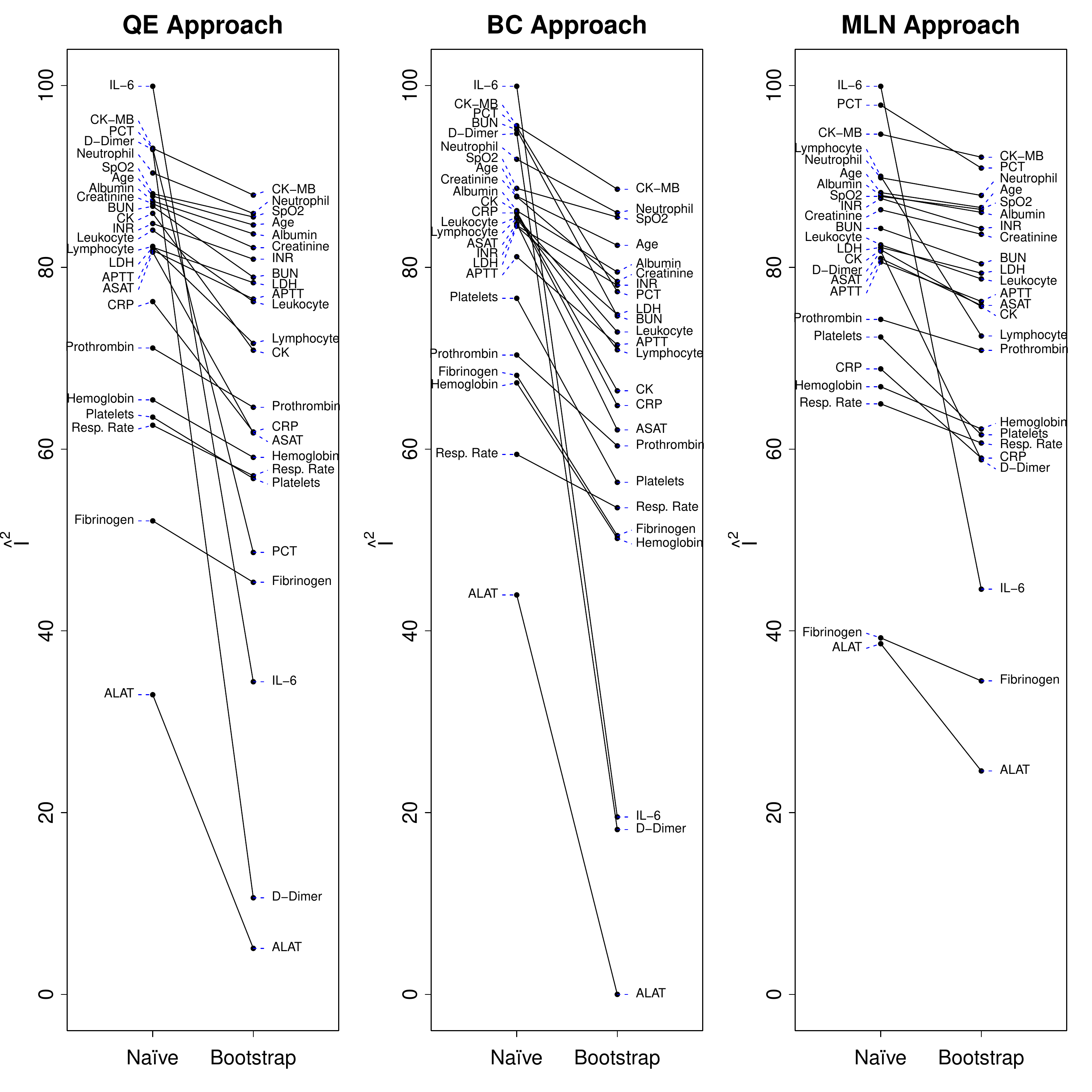}
    \caption{$\hat{I}^2$ values obtained when applying the naïve and bootstrap approaches to estimate the within-study SEs. SpO2 = Oxygen saturation; APTT = activated partial thrombin time; INR = International normalized ratio; ALAT = Alanine transaminase; ASAT = Aspartate transaminase; LDH = Lactate dehydrogenase; BUN = Blood urea nitrogen; CRP = C-reactive
protein; IL-6 = Interleukin-6; Resp Rate = Respiratory Rate; PCT = Procalcitonin CK = creatine kinase; CK-MB = creatine kinase–myocardial band. \label{fig:application}}
\end{figure}

The estimated pooled difference of means of all outcomes based on the MLN method are given in Table \ref{table:application} and those based on the QE and BC methods are given in Appendix \ref{sec: appendix application} for parsimony. Figure \ref{fig:application} illustrates the extent to which the $\hat{I}^2$ values changed when using within-study bootstrap SEs compared to the naïve SEs. The estimates of $\tau^2$ and the $I^2$ are listed in Appendix \ref{sec: appendix application}. 

For most outcomes, the pooled estimates and their 95\% CIs were very similar when using bootstrap estimates of the within-study SEs compared to the naïve estimates. The estimated between-study heterogeneity was nearly always smaller when using bootstrap estimates of the within-study SEs compared to the naïve estimates. The estimates of between-study variance decreased on average by 25\% for the QE method, 30\% for the BC method, and 13\% for the MLN method when using bootstrap SEs compared to when using the naïve SEs.  Moreover, the $\hat{I}^2$ values decreased on average by 15 percentage points for the QE method, 19 percentage points for the BC method, and 8 percentage points for the MLN method when using bootstrap SEs compared to when using the naïve SEs. 

The use of bootstrap SEs was most consequential in the analyses of the Interleukin-6 (IL-6) outcome. The estimates of the pooled difference of mean IL-6 (in pg/mL) between patients who died and patients who survived were 47.92 [13.66, 82.18], 41.73 [7.82, 75.64], and 40.64 [7.23, 74.05] when using the QE, BC, and MLN approaches, respectively, with naïve SEs. However, when using bootstrap SEs, the estimates of the pooled difference of mean IL-6 were 4.56 [3.13, 5.99], 4.34 [3.09, 5.60], and 4.77 [3.26, 6.28], respectively. The reason for this large discrepancy is that studies with large differences of mean IL-6 values had considerably smaller weights when using bootstrap SEs (Table \ref{table:application studylevel}). Estimation of heterogeneity was also strongly affected when using bootstrap SEs. The estimates of between-study variance were greater than 2,000 for the three transformation-based approaches when using the naïve SEs and were close to 1 when using bootstrap SEs. Similarly, the $\hat{I}^2$ values were greater than 99\% for these three transformation-based approaches when using the naïve SEs and were 34\%, 20\%, and 45\%, respectively, when using bootstrap SEs. These results may be attributed to the study-specific SEs being generally larger when using the bootstrap approach compared to the naïve approach.

\begin{table}[H] 
\caption{Data application results for the MLN approach. The entries in ``$S_2$ Studies" column are the number (percentage) of primary studies that reported $S_2$ summary data in both groups of patients (i.e., survivors and non-survivors).  \label{table:application}}
\centering
\begin{tabular}{@{\extracolsep{6pt}}lllll@{}}
\hline
& & & \multicolumn{2}{c}{Pooled Difference of Means Estimate [95\% CI]}\\ \cline{4-5}
Outcome & Studies & $S_2$ Studies & Naïve & Bootstrap\ \\
  \hline
  \textbf{Demographics} &  \\ 
\,\,\,\, Age (years) &  51 & 29 (57\%) & 12.82 [11.19, 14.45] & 12.83 [11.2, 14.46] \\ 
\textbf{Clinical Values} &  \\ 
\,\,\,\, SpO2-without O2 (\%) &  15 & 10 (67\%) & -6.91 [-8.73, -5.09] & -6.88 [-8.69, -5.07] \\  
\,\,\,\, Respiratory Rate (per min) &  13 & 10 (77\%) & 3.77 [2.81, 4.72] & 3.77 [2.8, 4.74] \\  
\textbf{Laboratory Values} & \\
\,\,\,\, Hemoglobin (g/L) &  18 & 12 (67\%) & -2.3 [-4.99, 0.39] & -2.26 [-4.92, 0.4] \\ 
\,\,\,\, Leukocyte ($10^9$/L) &  35 & 30 (86\%) & 3.16 [2.5, 3.82] & 3.14 [2.48, 3.81] \\ 
\,\,\,\, Lymphocyte ($10^9$/L) &  35 & 29 (83\%) & -0.37 [-0.43, -0.31] & -0.38 [-0.45, -0.32] \\ 
\,\,\,\, Neutrophil ($10^9$/L) &  22 & 19 (86\%) & 3.74 [2.86, 4.63] & 3.72 [2.83, 4.61] \\ 
\,\,\,\, Platelets ($10^9$/L) &  28 & 23 (82\%) & -34.64 [-44.95, -24.33] & -32.03 [-41.9, -22.17] \\ 
\,\,\,\, APTT (sec) &  14 & 9 (64\%) & 0.77 [-0.47, 2.01] & 0.76 [-0.43, 1.96] \\ 
\,\,\,\, D-Dimer (mg/L) &  19 & 18 (95\%) & 1.67 [0.95, 2.4] & 1.55 [0.87, 2.23] \\ 
\,\,\,\, Fibrinogen (g/L) &   7 & 4 (57\%) & 0.17 [-0.11, 0.45] & 0.18 [-0.1, 0.46] \\ 
\,\,\,\, INR &   6 & 5 (83\%) & 0.09 [0.02, 0.17] & 0.09 [0.02, 0.17] \\ 
\,\,\,\, Prothrombin (sec) &  24 & 18 (75\%) & 1.03 [0.78, 1.27] & 1.03 [0.78, 1.27] \\ 
\,\,\,\, ALAT (U/L) &  31 & 27 (87\%) & 4.3 [1.98, 6.62] & 4.13 [1.83, 6.44] \\ 
\,\,\,\, Albumin (g/L) &  19 & 14 (74\%) & -4.56 [-5.61, -3.51] & -4.56 [-5.62, -3.51] \\ 
\,\,\,\, ASAT (U/L) &  25 & 22 (88\%) & 16.74 [12.28, 21.2] & 16.46 [12.09, 20.83] \\ 
\,\,\,\, LDH (U/L) &  22 & 18 (82\%) & 210.58 [173.86, 247.3] & 209.02 [172.03, 246.01] \\
\,\,\,\, BUN (mmol/L) &  15 & 15 (100\%) & 3.47 [2.63, 4.31] & 3.44 [2.6, 4.27] \\ 
\,\,\,\, Creatinine ($\mu$mol/L) &  27 & 22 (81\%) & 21.74 [14.71, 28.77] & 21.42 [14.43, 28.41] \\ 
\,\,\,\, CRP (mg/L) &  26 & 19 (73\%) & 53.42 [40.47, 66.37] & 51.43 [38.17, 64.69] \\ 
\,\,\,\, IL-6 (pg/mL) &  10 & 10 (100\%) & 40.64 [7.23, 74.05] & 4.77 [3.26, 6.28] \\ 
\,\,\,\, PCT (ng/mL) &  10 & 8 (80\%) & 0.48 [0.25, 0.71] & 0.48 [0.23, 0.73] \\ 
\,\,\,\, CK (U/L) &  15 & 13 (87\%) & 122.19 [68.4, 175.98] & 113.33 [60.94, 165.71] \\ 
\,\,\,\, CK-MB (U/L) &   7 & 7 (100\%) & 6.43 [1.42, 11.44] & 6.02 [1.24, 10.81] \\ 
   \hline
\end{tabular}
\caption*{\small SpO2 = Oxygen saturation; APTT = activated partial thrombin time; INR = International normalized ratio; ALAT = Alanine transaminase; ASAT = Aspartate transaminase; LDH = Lactate dehydrogenase; BUN = Blood urea nitrogen; CRP = C-reactive
protein; IL-6 = Interleukin-6; PCT = Procalcitonin CK = creatine kinase; CK-MB = creatine kinase–myocardial band.}
\end{table}

\begin{table}[H] 
\caption{Study-specific estimates of the difference of means of Interleukin-6 (IL-6) using the MLN approach. All 10 primary studies reported the median IL-6 value in both groups. The unit of measurement for IL-6 is pg/mL. The number of patients who died is denoted by $n_1$ and the number of patients who survived is denoted by $n_2$. The columns under the ``Estimated SE" heading give the estimated within-study SEs of the difference of means based on the naïve and bootstrap approaches. The columns under the ``Weight (\%)" heading give the weights of the primary studies when using the naïve and bootstrap within-study SE estimators. \label{table:application studylevel}}
\centering
\begin{tabular}{@{\extracolsep{6pt}}llllllll@{}}
  \hline
& & & & \multicolumn{2}{c}{Estimated SE} & \multicolumn{2}{c}{Weight (\%)}\\    \cline{5-6} \cline{7-8}
Study & $n_1$ & $n_2$ &  Difference of Means & Naïve & Bootstrap & Naïve & Bootstrap \\ 
  \hline
Wang et al.\ & 65 & 274 & 155.12 & 35.46 & 49.52 & 8.48 & 0.02\\ 
Wu et al.\ & 44 &  40 & 5.46 & 0.97 & 1.02 & 13.38 & 21.92 \\ 
Li et al.\ & 15 &  87  &  142.37 & 116.86 & 106.50 & 1.84 & 0.01 \\ 
Chen et al.\ & 113  & 161  &  100.76 & 15.49 & 20.95 & 12.05 & 0.13 \\ 
Tu et al.\ & 25 & 149 &  16.65 & 73.17 & 180.89 & 3.86 & 0.00 \\ 
Sun et al.\ & 121 & 123 &  80.47 & 39.34 & 37.64 & 7.81 & 0.04 \\ 
Chen et al.\ & 81 & 337 &  2.63 & 0.55 & 0.63 & 13.38 & 28.76 \\ 
Zhou et al.\ & 54 & 137 &  5.20 & 0.82 & 0.90 & 13.38 & 23.93 \\ 
Fan et al.\ & 47 &  26 &  5.18 & 0.79 & 0.85 & 13.38 & 24.93 \\ 
Xu et al.\ & 28 & 117 &  26.82 & 12.79 & 15.25 & 12.45 & 0.25 \\ 
   \hline
\end{tabular}
\end{table}

\section{Discussion} \label{sec:discussion}

In this paper, we illustrated that the standard application of transformation-based approaches in meta-analysis can severely underestimate within-study SEs, which in turn can result in poor inference at the meta-analytic level. We proposed a straightforward application of parametric bootstrap to estimate within-study SEs in this context. Focusing on the recently proposed transformation-based approaches of McGrath et al.\ \cite{mcgrath2020estimating} and Cai et al.\ \cite{cai2021estimating}, we showed in a simulation study that (i) the bootstrap SE estimators often perform considerably better than the naïve SE estimators and (ii) inference on between-study heterogeneity improved when using the study-specific bootstrap SE estimators. The data application illustrated the extent to which conclusions may change when using the bootstrap SE estimators in real data applications. For instance, if one were to use the naïve SEs in our data application, one may have questioned whether it makes sense to meta-analyze some of the outcomes (e.g., Interleukin-6) due to the extremely large heterogeneity. However, when using the bootstrap SEs in subsequent analyses, we found that the large degree of heterogeneity was likely an artifact of the bias of the naïve SE estimators.

The performance of the pooled mean estimators was not strongly affected by whether the naïve or bootstrap within-study SE estimator was used, which may be expected. The bias of the pooled mean estimators was not strongly affect since the pooled mean estimator is unbiased for any choice of study-specific weights under the assumptions of the meta-analytic model. Moreover, the coverage of the 95\% CIs for the pooled mean was not strongly affected since the CIs depend on the total variance of the study-specific means.

While we focused on applying the proposed bootstrap SE estimators to meta-analyze the mean of an outcome of interest and meta-analyze the difference of means an outcome across two groups, the bootstrap SE estimators are applicable in much more general settings. In particular, these approaches can be directly applied for other effect measures (e.g., the standardized difference of means, ratio of means) and in more general meta-analytic settings (e.g., multivariate meta-analysis, network meta-analysis, multi-level meta-analysis, meta-regression). Exploring the impact of using the proposed parametric bootstrap SE estimators in these more complex settings would be an interesting area of further research. 

We considered some variations of the parametric bootstrap SE estimators in preliminary analyses. For instance, we considered estimating the standard deviation of the distribution of the bootstrap replicates based on the median absolute deviation (with appropriate scaling to ensure consistency under normality) instead of the sample standard deviation to gain robustness to outliers. Since these modifications did not clearly improve the performance of the estimator in our simulations, we did not include them for parsimony.

The proposed parametric bootstrap SE estimators are of course applicable to other transformation-based approaches, which may be another interesting avenue of further research. We focused on the methods of McGrath et al.\ \cite{mcgrath2020estimating} and Cai et al.\ \cite{cai2021estimating} because they have been shown to perform well for estimating the mean and standard deviation when data are skewed, which is often the case when primary studies report medians. For simpler transformation-based approaches whose mean and standard deviation estimators are linear combinations of sample quantiles, the SE can be analytically derived under parametric assumptions (e.g., see \cite{yang2021generalized}).

This study has a few limitations that are important to consider. First, the proposed bootstrap SE estimators may perform poorly when applied to primary studies with small sample sizes (e.g., less than 50). Second, since the proposed bootstrap SE estimators make parametric assumptions, one may expect some degree of model misspecification in practice. While the bootstrap SE estimators often performed reasonably well under model misspecification in our simulations, results may vary in different settings. 

Owing to their broad applicability, transformation-based approaches are most commonly applied when a meta-analysis includes primary studies that report the median of the outcome. However, it should be noted that a number of other approaches have been proposed in more case-specific contexts. One line of approaches estimates a pooled median or the difference of medians across two groups \cite{mcgrath2019one, mcgrath2020meta, ozturk2020meta}. Another approach estimates a global location parameter or the difference in global location parameters under a suitable location-scale model \cite{yang2021generalized}.

In summary, we recommend using the proposed bootstrap SE estimator when applying transformation-based approaches to meta-analyze studies reporting medians. While our results suggest that the MLN approach may be the preferred transformation-based approach when data are suspected to be non-normal, we encourage data analysts to perform sensitivity analyses evaluating the extent to which conclusions differ when applying other transformation-based approaches. To facilitate their application, we implemented the QE, BC, and MLN approaches with parametric bootstrap SEs in the \texttt{estmeansd} R package available on the Comprehensive R Archive Network (CRAN) \cite{estmeansd} and in the Shiny application available at \url{https://smcgrath.shinyapps.io/estmeansd/}.

\section*{Acknowledgements}
The authors thank Claudia Denkinger for helping collect the data set used in the data application and Siyu Cai for providing code implementing the MLN approach. The simulations in this work were run on the FASRC Cannon cluster supported by the FAS Division of Science Research Computing Group at Harvard University. This work was supported by the National Science Foundation Graduate Research Fellowship Program under Grant No. DGE1745303. Any opinions, findings, and conclusions or recommendations expressed in this material are those of the authors and do not necessarily reflect the views of the funding agencies.

\printbibliography

\newpage 

\appendix

\section{Additional results for the illustrative example} \label{sec: appendix illustrative example}

In this section, we give additional simulation results for the illustrative example in Section \ref{sec:naive SE}.

\subsection{Naïve standard errors when applying other transformation-based approaches}

We applied the transformation-based approaches suggested by Hozo et al.\ \cite{hozo2005estimating}, Wan et al.\ \cite{wan2014estimating}, Luo et al.\ \cite{luo2018optimally}, Shi et al.\ \cite{shi2020estimating}, and Yang et al.\ \cite{yang2021generalized} to the illustrative example. The approach of Yang et al.\ \cite{yang2021generalized} requires specifying a location-scale distribution for the outcome. We applied the approach of Yang et al.\ \cite{yang2021generalized} under the assumption of normality, as implemented in the \texttt{metaBLUE} R package \cite{metaBLUE}.

\begin{figure} [H]
    \centering
    \includegraphics[width=0.75\textwidth]{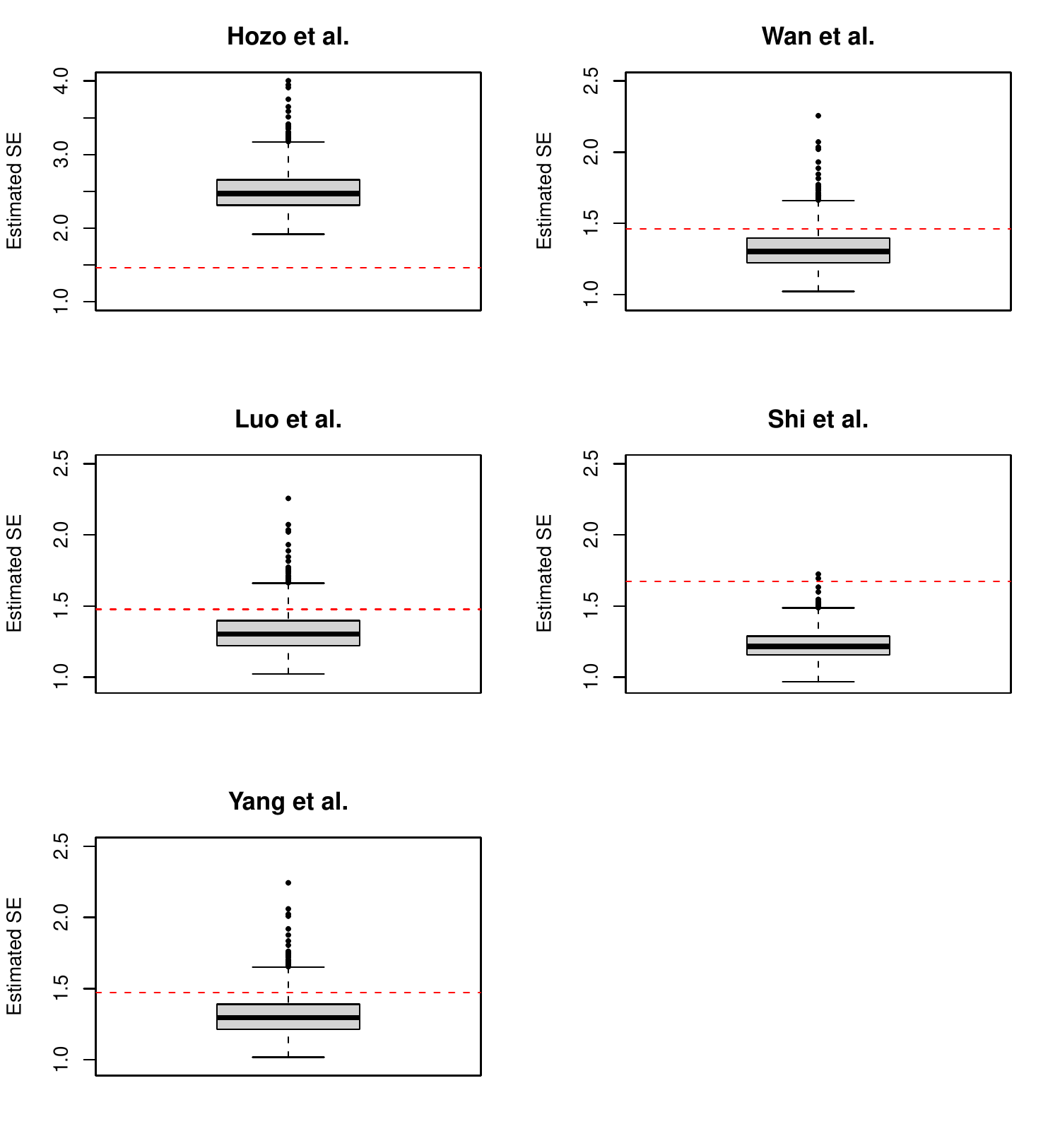}
    \caption{Boxplot of the naïve SE estimates of the approaches suggested by Hozo et al.\ \cite{hozo2005estimating}, Wan et al.\ \cite{wan2014estimating}, Luo et al.\ \cite{luo2018optimally}, Shi et al.\ \cite{shi2020estimating}, and Yang et al.\ \cite{yang2021generalized} from the example in Section \ref{sec:naive SE}. The dotted red lines illustrate the true SEs of the approaches. \label{fig:illustration other methods}}
\end{figure}

Figure \ref{fig:illustration other methods} gives the distribution of the naïve SE estimates along with the true SEs when applying these transformation-based approaches. As one may expect, the naïve SE estimator underestimated the true SE when applying the approaches of Wan et al.\ \cite{wan2014estimating}, Luo et al.\ \cite{luo2018optimally}, Shi et al.\ \cite{shi2020estimating}, and Yang et al.\ \cite{yang2021generalized}. However, naïve SE estimator overestimated the true SE when applying the approach of Hozo et al.\ \cite{hozo2005estimating}. This is simply due to the approach of Hozo et al.\ \cite{hozo2005estimating} severely overestimating the standard deviation of the underlying distribution. This can be seen more explicitly by noting that the standard deviation of the underlying distribution divided by $\sqrt{n}$ (i.e., the best-case scenario for this approach) is approximately 1.2, which is smaller than the true SE of the approach.

\subsection{Parametric bootstrap standard errors when applying the QE, BC, and MLN approaches}

\begin{figure} [H]
    \centering
    \includegraphics[width=\textwidth]{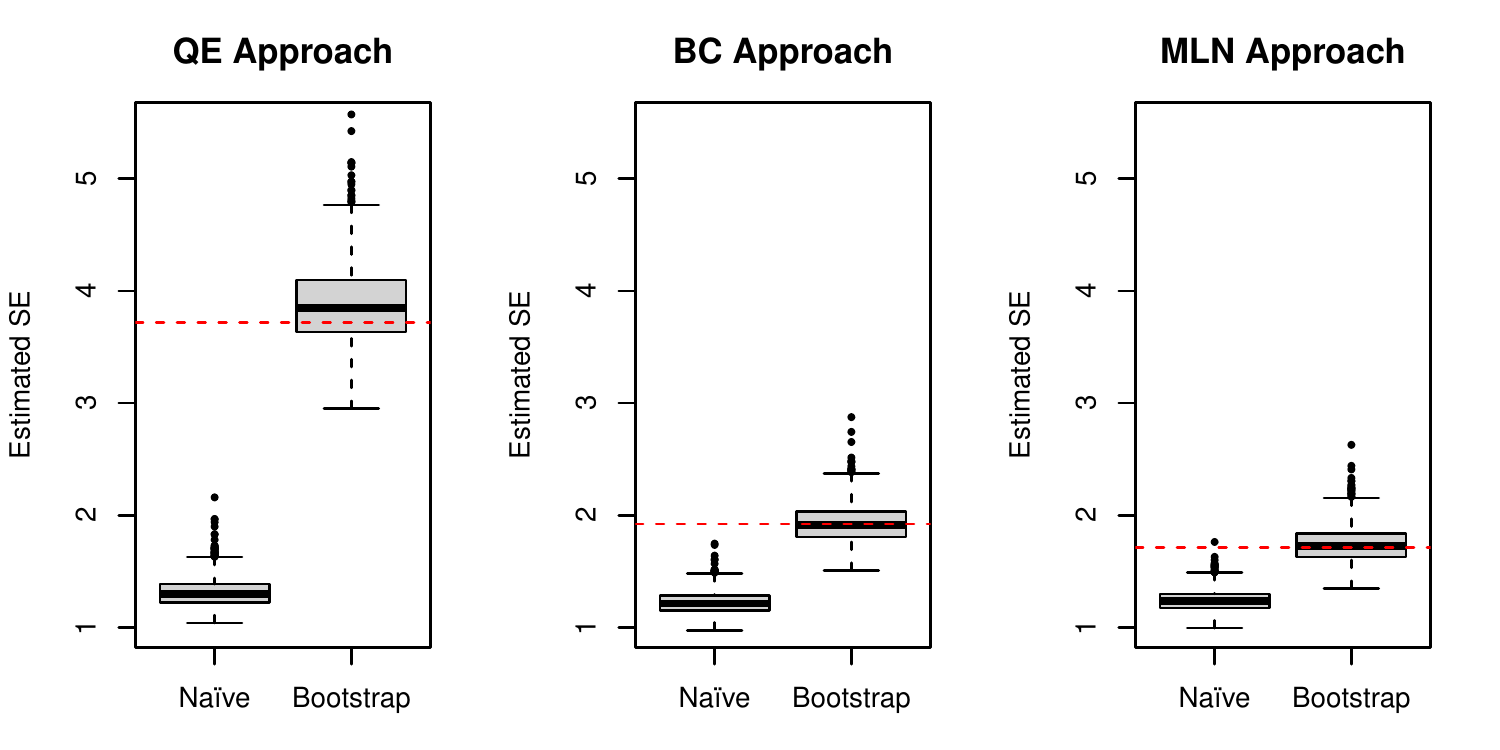}
    \caption{Boxplot of the naïve and bootstrap SE estimates of the QE, BC, and MLN approaches from the example in Section \ref{sec:naive SE}. The dotted red lines illustrate the true SEs of the three approaches. \label{fig:illustration with boot}}
\end{figure}

Figure \ref{fig:illustration with boot} includes the distributions of bootstrap SE estimates of the QE, BC, and MLN approaches.

\newpage
\section{Additional results for the study-level simulations} \label{sec: appendix study level}

\begin{figure} [H]
    \centering
    \includegraphics[width=\textwidth]{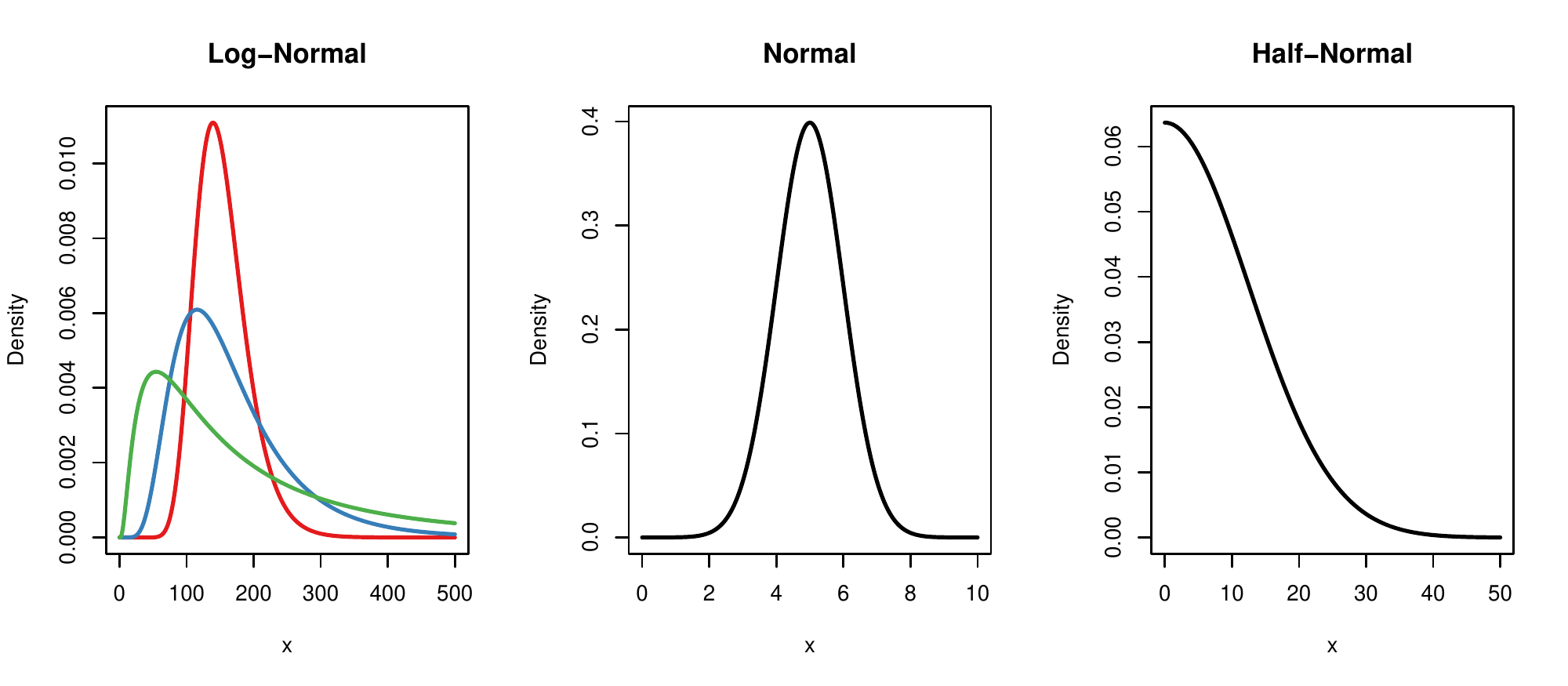}
    \caption{Probability density functions of the distributions included in the study-level simulations. In the left panel, the red line corresponds to the $\text{Log-Normal}(5, 0.25^2)$ distribution, the blue line corresponds to the $\text{Log-Normal}(5, 0.5^2)$ distribution, and the green line corresponds to the $\text{Log-Normal}(5, 1)$ distribution. The middle panel illustrates the $\text{Normal}(5, 1)$ distribution. The right panel illustrates the half-normal distribution with a mean of 10.  \label{fig:distributions}}
\end{figure}

Figure \ref{fig:distributions} illustrates the probability density functions of the distributions included in the simulations. Tables \ref{table:studylevel are S1}, \ref{table:studylevel are S2}, and \ref{table:studylevel are S3} give the study-level simulation results for the mean percent error. Tables \ref{table:studylevel rmse S1}, \ref{table:studylevel rmse S2}, and \ref{table:studylevel rmse S3} give the study-level simulation results for the RMSE.

\begin{table}[H]
\caption{Mean percent error of the naïve and parametric bootstrap SE estimators in scenario $S_1$. The mean percent error values are rounded to the nearest integer.\label{table:studylevel are S1}}
\begin{center}
\begin{tabular}{@{\extracolsep{6pt}}llllllll@{}}
\hline
& & \multicolumn{2}{c}{QE Approach} & \multicolumn{2}{c}{BC Approach} & \multicolumn{2}{c}{MLN Approach} \\ \cline{3-4} \cline{5-6} \cline{7-8} 
 Distribution & $n$ & Naïve & Boot.\ & Naïve & Boot. & Naïve  & Boot.\ \\
  \hline
$\text{Log-Normal}(5, 0.25^2)$  & 50 &  -20 & 10 & -17 & -2 & -15 & 0 \\ 
& 250 & -46 & 3 & -27 & -1 & -21 & 1 \\ 
& 1000 & -65 & 4 & -36 & 0 & -27 & 2 \\ 
$\text{Log-Normal}(5, 0.5^2)$ & 50 & -19 & 23 & -21 & -3 & -16 & 0 \\ 
& 250 & -40 & 12 & -32 & -2 & -27 & 2 \\ 
& 1000 & -58 & 10 & -45 & -1 & -39 & 2 \\ 
$\text{Log-Normal}(5, 1)$ & 50 & -20 & 202 & -24 & 1 & -15 & 8 \\ 
& 250 & -50 & 45 & -42 & -2 & -36 & 4 \\ 
& 1000 & -66 & 32 & -57 & 1 & -52 & 7 \\ 
$\text{Normal}(5, 1)$ & 50 & -18 & 12 & -16 & -1 & -10 & 5 \\ 
& 250 & -44 & 12 & -25 & 0 & -20 & 2 \\ 
& 1000 & -70 & 4 & -35 & 0 & -29 & 2 \\ 
$\text{Half-Normal}(0.1)$ & 50 & -15 & 13 & -25 & -5 & -15 & 9 \\ 
& 250 & -31 & 8 & -36 & -5 & -31 & 2 \\ 
& 1000 & -45 & 11 & -46 & 0 & -42 & 7 \\ 
   \hline
\end{tabular}
\end{center}
\end{table}

\begin{table}[H]
\caption{Mean percent error of the naïve and parametric bootstrap SE estimators in scenario $S_2$. The mean percent error values are rounded to the nearest integer.\label{table:studylevel are S2}}
\begin{center}
\begin{tabular}{@{\extracolsep{6pt}}llllllll@{}}
\hline
& & \multicolumn{2}{c}{QE Approach} & \multicolumn{2}{c}{BC Approach} & \multicolumn{2}{c}{MLN Approach} \\ \cline{3-4} \cline{5-6} \cline{7-8} 
 Distribution & $n$ & Naïve & Boot.\ & Naïve & Boot. & Naïve  & Boot.\ \\
  \hline
$\text{Log-Normal}(5, 0.25^2)$  & 50 &  -12 & -3 & -19 & 0 & -13 & -5 \\ 
& 250 & -19 & -1 & -29 & 1 & -11 & -1 \\ 
& 1000 & -27 & 2 & -35 & 5 & -13 & -1 \\ 
$\text{Log-Normal}(5, 0.5^2)$ & 50 & -18 & -4 & -27 & -5 & -16 & -7 \\ 
& 250 & -22 & -4 & -33 & -1 & -15 & -2 \\ 
& 1000 & -30 & -1 & -38 & 2 & -16 & 0 \\ 
$\text{Log-Normal}(5, 1)$ & 50 & -28 & 15 & -36 & -2 & -18 & -5 \\ 
& 250 & -29 & 0 & -41 & -2 & -19 & -1 \\ 
& 1000 & -21 & 0 & -46 & 0 & -21 & -1 \\ 
$\text{Normal}(5, 1)$ & 50 & -12 & -2 & -18 & 2 & -9 & -2 \\ 
& 250 & -23 & -3 & -34 & -3 & -6 & 3 \\  
& 1000 & -37 & -7 & -46 & -4 & -11 & 1 \\ 
$\text{Half-Normal}(0.1)$ & 50 & -27 & 11 & -39 & 9 & 11 & 28 \\ 
& 250 & -46 & 2 & -53 & 1 & -28 & 4 \\ 
& 1000 & -51 & -6 & -59 & 1 & -56 & 9 \\ 
   \hline
\end{tabular}
\end{center}
\end{table}

\begin{table}[H]
\caption{Mean percent error of the naïve and parametric bootstrap SE estimators in scenario $S_3$. The mean percent error values are rounded to the nearest integer.\label{table:studylevel are S3}}
\begin{center}
\begin{tabular}{@{\extracolsep{6pt}}llllllll@{}}
\hline
& & \multicolumn{2}{c}{QE Approach} & \multicolumn{2}{c}{BC Approach} & \multicolumn{2}{c}{MLN Approach} \\ \cline{3-4} \cline{5-6} \cline{7-8} 
 Distribution & $n$ & Naïve & Boot.\ & Naïve & Boot. & Naïve  & Boot.\ \\
  \hline
$\text{Log-Normal}(5, 0.25^2)$  & 50 &  -2 & 12 & -7 & 0 & -6 & -2 \\ 
& 250 & -22 & 5 & -15 & -1 & -8 & 0 \\ 
& 1000 & -42 & 4 & -25 & 0 & -12 & 0 \\ 
$\text{Log-Normal}(5, 0.5^2)$ & 50 & -1 & 27 & -11 & -3 & -8 & -3 \\ 
& 250 & -17 & 14 & -18 & -1 & -10 & -1 \\ 
& 1000 & -35 & 11 & -30 & -1 & -15 & -1 \\ 
$\text{Log-Normal}(5, 1)$ & 50 & -10 & 799 & -12 & -1 & -4 & -1 \\ 
& 250 & -34 & 62 & -22 & -3 & -10 & -1 \\ 
& 1000 & -39 & 49 & -38 & -2 & -18 & -1 \\ 
$\text{Normal}(5, 1)$ & 50 & -2 & 9 & -6 & 1 & -2 & 3 \\ 
& 250 & -20 & 9 & -15 & 0 & -8 & -1 \\ 
& 1000 & -50 & 4 & -25 & 1 & -16 & -1 \\ 
$\text{Half-Normal}(0.1)$ & 50 & -2 & 14 & -9 & 36 & 7 & 21 \\ 
& 250 & -15 & -1 & -19 & 3 & -7 & 12 \\
& 1000 & -22 & 2 & -30 & 4 & -13 & 19 \\ 
   \hline
\end{tabular}
\end{center}
\end{table}

\begin{table}[H]
\caption{Root mean squared error of the naïve and parametric bootstrap SE estimators in scenario $S_1$. \label{table:studylevel rmse S1}}
\begin{center}
\begin{tabular}{@{\extracolsep{6pt}}llllllll@{}}
\hline
& & \multicolumn{2}{c}{QE Approach} & \multicolumn{2}{c}{BC Approach} & \multicolumn{2}{c}{MLN Approach} \\ \cline{3-4} \cline{5-6} \cline{7-8} 
 Distribution & $n$ & Naïve & Boot.\ & Naïve & Boot. & Naïve  & Boot.\ \\
  \hline
$\text{Log-Normal}(5, 0.25^2)$  & 50 &  1.98 & 1.98 & 1.45 & 1.10 & 1.29 & 1.05 \\ 
& 250 & 2.30 & 0.73 & 0.92 & 0.37 & 0.71 & 0.36 \\ 
& 1000 & 2.40 & 0.39 & 0.70 & 0.18 & 0.48 & 0.17 \\ 
$\text{Log-Normal}(5, 0.5^2)$ & 50 & 5.83 & 10.86 & 4.30 & 3.66 & 3.69 & 3.44 \\ 
& 250 & 4.34 & 3.18 & 2.79 & 1.32 & 2.25 & 1.28 \\ 
& 1000 & 4.24 & 1.75 & 2.35 & 0.68 & 1.89 & 0.67 \\ 
$\text{Log-Normal}(5, 1)$ & 50 & 149.52 & 947.67 & 26.35 & 31.16 & 24.50 & 30.14 \\ 
& 250 & 29.90 & 60.94 & 15.31 & 9.61 & 12.91 & 9.66 \\ 
& 1000 & 25.05 & 33.70 & 13.17 & 5.34 & 11.66 & 5.48 \\ 
$\text{Normal}(5, 1)$ & 50 & 0.04 & 0.04 & 0.03 & 0.02 & 0.03 & 0.03 \\ 
& 250 & 0.05 & 0.03 & 0.02 & 0.01 & 0.02 & 0.01 \\ 
& 1000 & 0.08 & 0.02 & 0.02 & 0.00 & 0.01 & 0.00 \\ 
$\text{Half-Normal}(0.1)$ & 50 & 0.28 & 0.35 & 0.36 & 0.23 & 0.28 & 0.30 \\ 
& 250 & 0.21 & 0.12 & 0.24 & 0.09 & 0.20 & 0.09 \\ 
& 1000 & 0.17 & 0.06 & 0.18 & 0.04 & 0.15 & 0.05 \\ 
   \hline
\end{tabular}
\end{center}
\end{table}

\begin{table}[H]
\caption{Root mean squared error of the naïve and parametric bootstrap SE estimators in scenario $S_2$. \label{table:studylevel rmse S2}}
\begin{center}
\begin{tabular}{@{\extracolsep{6pt}}llllllll@{}}
\hline
& & \multicolumn{2}{c}{QE Approach} & \multicolumn{2}{c}{BC Approach} & \multicolumn{2}{c}{MLN Approach} \\ \cline{3-4} \cline{5-6} \cline{7-8} 
 Distribution & $n$ & Naïve & Boot.\ & Naïve & Boot. & Naïve  & Boot.\ \\
  \hline
$\text{Log-Normal}(5, 0.25^2)$  & 50 &  1.21 & 1.15 & 1.60 & 1.13 & 1.19 & 1.08 \\ 
& 250 & 0.59 & 0.29 & 1.01 & 0.29 & 0.35 & 0.24 \\ 
& 1000 & 0.45 & 0.12 & 0.66 & 0.16 & 0.19 & 0.07 \\ 
$\text{Log-Normal}(5, 0.5^2)$ & 50 & 4.12 & 3.84 & 5.41 & 4.05 & 3.60 & 3.39 \\ 
& 250 & 1.71 & 1.00 & 2.75 & 0.95 & 1.15 & 0.79 \\ 
& 1000 & 1.21 & 0.29 & 1.68 & 0.31 & 0.56 & 0.21 \\ 
$\text{Log-Normal}(5, 1)$ & 50 & 38.93 & 102.88 & 34.11 & 35.19 & 24.04 & 24.35 \\ 
& 250 & 9.98 & 6.91 & 13.52 & 6.65 & 6.30 & 5.09 \\ 
& 1000 & 2.96 & 1.42 & 7.92 & 1.85 & 2.92 & 1.32 \\ 
$\text{Normal}(5, 1)$ & 50 & 0.03 & 0.03 & 0.04 & 0.03 & 0.03 & 0.03 \\ 
& 250 & 0.02 & 0.01 & 0.03 & 0.01 & 0.01 & 0.01 \\ 
& 1000 & 0.02 & 0.01 & 0.03 & 0.00 & 0.00 & 0.00 \\ 
$\text{Half-Normal}(0.1)$ & 50 & 0.76 & 1.04 & 0.97 & 0.95 & 0.90 & 1.13 \\ 
& 250 & 0.49 & 0.27 & 0.55 & 0.26 & 0.34 & 0.20 \\ 
& 1000 & 0.28 & 0.09 & 0.32 & 0.09 & 0.33 & 0.08 \\ 
   \hline
\end{tabular}
\end{center}
\end{table}

\begin{table}[H]
\caption{Root mean squared error of the naïve and parametric bootstrap SE estimators in scenario $S_3$. \label{table:studylevel rmse S3}}
\begin{center}
\begin{tabular}{@{\extracolsep{6pt}}llllllll@{}}
\hline
& & \multicolumn{2}{c}{QE Approach} & \multicolumn{2}{c}{BC Approach} & \multicolumn{2}{c}{MLN Approach} \\ \cline{3-4} \cline{5-6} \cline{7-8} 
 Distribution & $n$ & Naïve & Boot.\ & Naïve & Boot. & Naïve  & Boot.\ \\
  \hline
$\text{Log-Normal}(5, 0.25^2)$  & 50 &  1.18 & 1.57 & 0.86 & 0.81 & 0.80 & 0.77 \\ 
& 250 & 0.84 & 0.53 & 0.47 & 0.21 & 0.27 & 0.19 \\ 
& 1000 & 0.97 & 0.23 & 0.42 & 0.09 & 0.17 & 0.06 \\ 
$\text{Log-Normal}(5, 0.5^2)$ & 50 & 4.93 & 10.86 & 2.73 & 2.51 & 2.36 & 2.27 \\ 
& 250 & 1.76 & 2.35 & 1.36 & 0.70 & 0.81 & 0.59 \\ 
& 1000 & 1.69 & 1.10 & 1.20 & 0.29 & 0.50 & 0.18 \\ 
$\text{Log-Normal}(5, 1)$ & 50 & 122.89 & 9101.26 & 18.30 & 18.15 & 15.94 & 15.65 \\ 
& 250 & 19.84 & 67.31 & 7.20 & 4.66 & 4.35 & 3.78 \\ 
& 1000 & 8.76 & 24.51 & 6.06 & 1.86 & 2.37 & 1.10 \\ 
$\text{Normal}(5, 1)$ & 50 & 0.02 & 0.03 & 0.02 & 0.02 & 0.02 & 0.02 \\ 
& 250 & 0.02 & 0.01 & 0.01 & 0.00 & 0.01 & 0.00 \\ 
& 1000 & 0.03 & 0.01 & 0.01 & 0.00 & 0.01 & 0.00 \\ 
$\text{Half-Normal}(0.1)$ & 50 & 0.21 & 0.43 & 0.50 & 3.12 & 0.26 & 0.37 \\ 
& 250 & 0.09 & 0.06 & 0.12 & 0.07 & 0.06 & 0.09 \\ 
& 1000 & 0.07 & 0.03 & 0.10 & 0.03 & 0.04 & 0.06 \\ 
   \hline
\end{tabular}
\end{center}
\end{table}

\newpage
\section{Additional results for the meta-analysis simulations}\label{sec: appendix meta-analytic level}

Tables \ref{table:meta-analysis pooled mean bias}, \ref{table:meta-analysis pooled mean var}, and \ref{table:meta-analysis pooled mean cov} gives the meta-analysis simulation results for the bias, variance, and coverage of the 95\% CIs of the pooled mean estimators. Table \ref{table:meta-analysis tausq var} gives the meta-analysis simulation results for the variance of the $\tau^2$ estimators. Table \ref{table:meta-analysis I2 median bias} gives the median bias (i.e., median of $\hat{I}^2 - I^2$) of the $I^2$ estimators

\begin{table}[H]
\caption{Bias of the pooled mean estimators. The number of primary studies is denoted by $K$ and the proportion of primary studies reporting medians is denoted by $p$. The last two rows correspond to the scenario where all studies report the sample mean and standard deviation of the outcome, in which case the transformation-based approaches are not applicable. The true value of the pooled mean is approximately 153.12. \label{table:meta-analysis pooled mean bias}}
\begin{center}
\begin{tabular}{@{\extracolsep{6pt}}lllllllll@{}}
\hline
& & & \multicolumn{2}{c}{QE Approach} & \multicolumn{2}{c}{BC Approach} & \multicolumn{2}{c}{MLN Approach} \\ \cline{4-5} \cline{6-7} \cline{8-9} 
Scenario & $K$ & $p$ & Naïve & Boot. & Naïve & Boot. &  Naïve & Boot. \\
  \hline
$S_1$ & 9 & $1/3$   &  0.50 & 0.12 & -0.16 & -0.18 & -0.08 & -0.11 \\ 
& & $2/3$   & 1.18 & 0.62 & -0.24 & -0.25 & -0.04 & -0.07 \\ 
& & $1$   & 1.85 & 1.44 & -0.30 & -0.36 & -0.02 & -0.06 \\ 
 & 30 & $1/3$   &  0.58 & 0.21 & -0.11 & -0.11 & -0.01 & -0.03 \\ 
& & $2/3$   & 1.20 & 0.58 & -0.21 & -0.23 & -0.02 & -0.05 \\ 
& & $1$   & 1.84 & 1.34 & -0.33 & -0.40 & -0.06 & -0.11 \\ 
$S_2$ & 9 & $1/3$   &  -0.41 & -0.37 & -0.48 & -0.38 & -0.14 & -0.13 \\ 
& & $2/3$   & -0.75 & -0.72 & -0.91 & -0.77 & -0.16 & -0.16 \\ 
& & $1$   & -1.06 & -1.09 & -1.28 & -1.25 & -0.20 & -0.21 \\ 
 & 30 & $1/3$   &  -0.40 & -0.37 & -0.48 & -0.37 & -0.11 & -0.11 \\ 
& & $2/3$   & -0.70 & -0.67 & -0.85 & -0.73 & -0.12 & -0.13 \\ 
& & $1$   & -1.08 & -1.11 & -1.30 & -1.28 & -0.22 & -0.23 \\ 
$S_3$ & 9 & $1/3$   &  0.18 & 0.08 & -0.18 & -0.17 & -0.13 & -0.13 \\ 
& & $2/3$   & 0.61 & 0.44 & -0.15 & -0.15 & -0.06 & -0.06 \\ 
& & $1$   & 0.89 & 0.75 & -0.27 & -0.28 & -0.13 & -0.14 \\ 
& 30 & $1/3$   &  0.26 & 0.15 & -0.11 & -0.11 & -0.08 & -0.08 \\ 
& & $2/3$   & 0.57 & 0.39 & -0.17 & -0.17 & -0.09 & -0.10 \\ 
& & $1$   & 0.91 & 0.74 & -0.25 & -0.26 & -0.12 & -0.13 \\  \hdashline 
& 9 & $0$ & -0.06 & -0.06 & -0.06 & -0.06 & -0.06 & -0.06 \\  
& 30 & $0$ & -0.05 & -0.05 & -0.05 & -0.05 & -0.05 & -0.05 \\ 
   \hline
\end{tabular}
\end{center}
\end{table}

\begin{table}[H]
\caption{Variance of the pooled mean estimators. The number of primary studies is denoted by $K$ and the proportion of primary studies reporting medians is denoted by $p$. The last two rows correspond to the scenario where all studies report the sample mean and standard deviation of the outcome, in which case the transformation-based approaches are not applicable.  \label{table:meta-analysis pooled mean var}}
\begin{center}
\begin{tabular}{@{\extracolsep{6pt}}lllllllll@{}}
\hline
& & & \multicolumn{2}{c}{QE Approach} & \multicolumn{2}{c}{BC Approach} & \multicolumn{2}{c}{MLN Approach} \\ \cline{4-5} \cline{6-7} \cline{8-9} 
Scenario & $K$ & $p$ & Naïve & Boot. & Naïve & Boot. &  Naïve & Boot. \\
  \hline
$S_1$ & 9 & $1/3$   &  2.06 & 1.76 & 1.62 & 1.59 & 1.52 & 1.52 \\ 
& & $2/3$   & 2.51 & 2.14 & 1.69 & 1.66 & 1.59 & 1.58 \\ 
& & $1$   & 3.32 & 3.19 & 1.90 & 1.91 & 1.68 & 1.69 \\ 
 & 30 & $1/3$   &  0.60 & 0.52 & 0.46 & 0.45 & 0.44 & 0.44 \\ 
& & $2/3$   & 0.72 & 0.60 & 0.45 & 0.44 & 0.43 & 0.42 \\ 
& & $1$   & 1.13 & 1.05 & 0.58 & 0.58 & 0.53 & 0.53 \\ 
$S_2$ & 9 & $1/3$   &  1.40 & 1.37 & 1.54 & 1.45 & 1.33 & 1.32 \\ 
& & $2/3$   & 1.44 & 1.44 & 1.61 & 1.61 & 1.35 & 1.36 \\ 
& & $1$   & 1.71 & 1.72 & 1.97 & 1.98 & 1.55 & 1.55 \\ 
 & 30 & $1/3$   &  0.41 & 0.41 & 0.43 & 0.42 & 0.39 & 0.39 \\ 
& & $2/3$   & 0.44 & 0.44 & 0.51 & 0.50 & 0.40 & 0.40 \\ 
& & $1$   & 0.48 & 0.49 & 0.57 & 0.57 & 0.43 & 0.44 \\ 
$S_3$ & 9 & $1/3$   &  1.51 & 1.50 & 1.39 & 1.39 & 1.36 & 1.36 \\ 
& & $2/3$   & 1.70 & 1.68 & 1.51 & 1.50 & 1.44 & 1.43 \\ 
& & $1$   & 2.00 & 1.98 & 1.54 & 1.54 & 1.45 & 1.45 \\ 
& 30 & $1/3$   &  0.40 & 0.40 & 0.38 & 0.38 & 0.37 & 0.37 \\ 
& & $2/3$   & 0.50 & 0.49 & 0.42 & 0.42 & 0.38 & 0.38 \\ 
& & $1$   & 0.61 & 0.60 & 0.47 & 0.48 & 0.42 & 0.42 \\  \hdashline 
& 9 & $0$ & 1.27 & 1.27 & 1.27 & 1.27 & 1.27 & 1.27 \\ 
& 30 & $0$ & 0.41 & 0.41 & 0.41 & 0.41 & 0.41 & 0.41 \\ 
\hline
\end{tabular}
\end{center}
\end{table}

\begin{table}[H]
\caption{Coverage of the 95\% confidence intervals for the pooled mean. The number of primary studies is denoted by $K$ and the proportion of primary studies reporting medians is denoted by $p$. The last two rows correspond to the scenario where all studies report the sample mean and standard deviation of the outcome, in which case the transformation-based approaches are not applicable. \label{table:meta-analysis pooled mean cov}}
\begin{center}
\begin{tabular}{@{\extracolsep{6pt}}lllllllll@{}}
\hline
& & & \multicolumn{2}{c}{QE Approach} & \multicolumn{2}{c}{BC Approach} & \multicolumn{2}{c}{MLN Approach} \\ \cline{4-5} \cline{6-7} \cline{8-9} 
Scenario & $K$ & $p$ & Naïve & Boot. & Naïve & Boot. &  Naïve & Boot. \\
  \hline
$S_1$ & 9 & $1/3$   &  0.91 & 0.91 & 0.90 & 0.90 & 0.91 & 0.91 \\ 
& & $2/3$   & 0.87 & 0.91 & 0.91 & 0.91 & 0.92 & 0.92 \\ 
& & $1$   & 0.83 & 0.88 & 0.90 & 0.91 & 0.92 & 0.93 \\ 
& 30 & $1/3$   &  0.88 & 0.92 & 0.93 & 0.92 & 0.93 & 0.92 \\ 
& & $2/3$   & 0.72 & 0.88 & 0.93 & 0.94 & 0.95 & 0.95 \\ 
& & $1$   & 0.51 & 0.72 & 0.90 & 0.89 & 0.92 & 0.92 \\ 
$S_2$ & 9 & $1/3$   &  0.92 & 0.92 & 0.90 & 0.92 & 0.91 & 0.92 \\ 
& & $2/3$   & 0.87 & 0.88 & 0.88 & 0.89 & 0.92 & 0.92 \\ 
& & $1$   & 0.82 & 0.83 & 0.80 & 0.83 & 0.90 & 0.91 \\ 
& 30 & $1/3$   & 0.89 & 0.90 & 0.88 & 0.90 & 0.93 & 0.93 \\ 
& & $2/3$   & 0.80 & 0.80 & 0.76 & 0.80 & 0.93 & 0.93 \\ 
& & $1$   & 0.64 & 0.62 & 0.57 & 0.59 & 0.93 & 0.93 \\ 
$S_3$ & 9 & $1/3$   &  0.91 & 0.91 & 0.91 & 0.91 & 0.91 & 0.91 \\ 
& & $2/3$   & 0.89 & 0.90 & 0.91 & 0.91 & 0.91 & 0.91 \\ 
& & $1$   & 0.85 & 0.89 & 0.90 & 0.91 & 0.90 & 0.91 \\ 
& 30 & $1/3$   &  0.94 & 0.94 & 0.94 & 0.94 & 0.94 & 0.94 \\ 
& & $2/3$   & 0.88 & 0.90 & 0.93 & 0.93 & 0.94 & 0.93 \\ 
& & $1$   & 0.75 & 0.81 & 0.91 & 0.91 & 0.93 & 0.93 \\ \hdashline 
& 9 & $0$ & 0.91 & 0.91 & 0.91 & 0.91 & 0.91 & 0.91 \\ 
& 30 & $0$ & 0.94 & 0.94 & 0.94 & 0.94 & 0.94 & 0.94 \\ 
\hline
\end{tabular}
\end{center}
\end{table}

\begin{table}[H]
\caption{Variance of the $\tau^2$ estimators. The number of primary studies is denoted by $K$ and the proportion of primary studies reporting medians is denoted by $p$. The last two rows correspond to the scenario where all studies report the sample mean and standard deviation of the outcome, in which case the transformation-based approaches are not applicable.  \label{table:meta-analysis tausq var}}
\begin{center}
\begin{tabular}{@{\extracolsep{6pt}}lllllllll@{}}
\hline
& & & \multicolumn{2}{c}{QE Approach} & \multicolumn{2}{c}{BC Approach} & \multicolumn{2}{c}{MLN Approach} \\ \cline{4-5} \cline{6-7} \cline{8-9} 
Scenario & $K$ & $p$ & Naïve & Boot. & Naïve & Boot. &  Naïve & Boot. \\
  \hline
$S_1$ & 9 & $1/3$   &  105.61 & 43.21 & 44.07 & 38.37 & 40.01 & 36.22 \\ 
& & $2/3$   & 177.25 & 73.70 & 55.00 & 46.55 & 48.39 & 42.11 \\ 
& & $1$   & 235.35 & 116.15 & 66.79 & 55.67 & 58.29 & 49.48 \\ 
 & 30 & $1/3$   &  29.33 & 12.08 & 12.25 & 10.77 & 11.25 & 10.28 \\ 
& & $2/3$   & 47.36 & 18.33 & 15.70 & 13.88 & 14.56 & 13.09 \\ 
& & $1$   & 59.40 & 26.71 & 15.82 & 14.35 & 13.50 & 12.41 \\ 
$S_2$ & 9 & $1/3$   &  37.69 & 35.04 & 46.61 & 38.24 & 31.59 & 30.67 \\ 
& & $2/3$   & 43.30 & 38.69 & 55.59 & 42.57 & 35.87 & 33.67 \\ 
& & $1$   & 43.65 & 38.80 & 63.90 & 48.63 & 36.77 & 34.23 \\ 
 & 30 & $1/3$   &  11.64 & 11.16 & 14.27 & 12.04 & 10.10 & 9.96 \\ 
& & $2/3$   & 12.24 & 11.67 & 17.55 & 14.50 & 9.72 & 9.37 \\ 
& & $1$   & 13.59 & 13.06 & 19.88 & 17.76 & 11.79 & 11.38 \\ 
$S_3$ & 9 & $1/3$   &  42.65 & 35.44 & 36.09 & 34.36 & 32.30 & 31.72 \\ 
& & $2/3$   & 64.32 & 50.29 & 42.49 & 40.28 & 37.73 & 36.64 \\ 
& & $1$   & 67.34 & 48.79 & 45.65 & 41.21 & 35.53 & 33.90 \\ 
& 30 & $1/3$   &  11.96 & 10.31 & 9.58 & 9.32 & 8.87 & 8.80 \\ 
& & $2/3$   & 15.34 & 12.80 & 11.24 & 10.95 & 9.40 & 9.30 \\ 
& & $1$   & 19.23 & 16.74 & 12.81 & 12.44 & 10.89 & 10.65 \\ \hdashline 
& 9 & $0$ & 29.35 & 29.35 & 29.35 & 29.35 & 29.35 & 29.35 \\ 
& 30 & $0$ & 8.86 & 8.86 & 8.86 & 8.86 & 8.86 & 8.86 \\ 
   \hline
\end{tabular}
\end{center}
\end{table}

\begin{table}[H]
\caption{Median bias of the $I^2$ estimators (i.e., median of $\hat{I}^2 - I^2$). The number of primary studies is denoted by $K$ and the proportion of primary studies reporting medians is denoted by $p$. The last two rows correspond to the scenario where all studies report the sample mean and standard deviation of the outcome, in which case the transformation-based approaches are not applicable. The true values of $I^2$ depend on the summary statistics reported, $p$, and the transformation-based approach. The true values of $I^2$ ranged from 19\% to 50\%. \label{table:meta-analysis I2 median bias}}
\begin{center}
\begin{tabular}{@{\extracolsep{6pt}}lllllllll@{}}
\hline
& & & \multicolumn{2}{c}{QE Approach} & \multicolumn{2}{c}{BC Approach} & \multicolumn{2}{c}{MLN Approach} \\ \cline{4-5} \cline{6-7} \cline{8-9} 
Scenario & $K$ & $p$ & Naïve & Boot. & Naïve & Boot. &  Naïve & Boot. \\
  \hline
$S_1$ & 9 & $1/3$   & 34.25 & 7.13 & 13.57 & 1.62 & 8.70 & -0.20 \\ 
& & $2/3$   & 49.22 & 1.63 & 21.96 & 0.20 & 16.53 & -1.33 \\ 
& & $1$   & 57.05 & -16.62 & 29.91 & -0.11 & 22.55 & -2.95 \\ 
 & 30 & $1/3$   &  35.94 & 14.46 & 15.78 & 6.48 & 12.29 & 4.73 \\ 
& & $2/3$   & 51.11 & 8.59 & 24.37 & 5.06 & 19.16 & 3.17 \\ 
& & $1$   & 58.74 & -11.07 & 31.60 & 2.20 & 24.44 & 0.45 \\ 
$S_2$ & 9 & $1/3$   &  8.29 & 1.04 & 14.52 & 2.33 & 1.27 & -2.11 \\ 
& & $2/3$   & 14.39 & 1.04 & 25.48 & 0.24 & 4.69 & -1.88 \\ 
& & $1$   & 18.70 & -2.22 & 31.41 & -7.27 & 7.91 & -2.15 \\ 
 & 30 & $1/3$   &  12.36 & 6.13 & 18.42 & 6.41 & 6.84 & 3.58 \\ 
& & $2/3$   & 17.94 & 5.30 & 27.99 & 4.73 & 9.72 & 3.06 \\ 
& & $1$   & 22.26 & 2.60 & 34.67 & 0.15 & 12.63 & 3.10 \\ 
$S_3$ & 9 & $1/3$   &  11.67 & -0.60 & 5.08 & -1.07 & 1.76 & -1.49 \\ 
& & $2/3$   & 23.06 & -1.02 & 11.55 & -0.65 & 5.51 & -0.53 \\ 
& & $1$   & 26.42 & -11.03 & 14.71 & -2.69 & 6.29 & -1.75 \\ 
& 30 & $1/3$   &  16.59 & 6.44 & 10.12 & 4.82 & 6.92 & 4.03 \\ 
& & $2/3$   & 24.66 & 3.86 & 15.06 & 4.24 & 9.06 & 3.79 \\ 
& & $1$   & 30.15 & -5.09 & 18.76 & 2.11 & 10.05 & 2.05 \\   \hdashline 
& 9 & $0$ & -0.85 & -0.85 & -0.85 & -0.85 & -0.85 & -0.85 \\ 
& 30 & $0$ & 2.85 & 2.85 & 2.85 & 2.85 & 2.85 & 2.85 \\ 
   \hline
\end{tabular}
\end{center}
\end{table}

\newpage
\section{Additional description and results for the data application} \label{sec: appendix application}

\subsection{Data processing}

Our analyses use the data sets from the sensitivity analyses of Katzenschlager et al.\ \cite{katzenschlager2021can} that removed two primary studies in the analysis of the C-reactive protein (CRP) outcome and one primary study in the analysis of the D-dimer outcome because they had large outlier values. Four additional data processing steps were performed for our analyses: 
\begin{enumerate}
    \item For each outcome, we removed primary studies that reported $S_2$ summary statistics with fewer than 10 individuals in any of the two comparison groups (i.e., survivors and non-survivors). These studies contribute little data and the transformation-based methods may perform poorly when applied to studies with such small sample sizes. For each outcome, this criterion removed at most three primary studies.
    \item Occasionally, some of the extracted sample quantiles of an outcome in a primary study equalled each other due to rounding. To avoid complications for the transformation-based approaches, we increased the value of the relevant quantile by 2.5\% to break such ties (e.g., we added a value of $0.025q_3$ to $q_3$ if $q_2 = q_3$). This occurred in seven primary studies for the respiratory rate outcome, two primary studies for the international normalized ratio (INR) outcome, and one primary study for the fibrinogen outcome; this did not occur in any primary study for the other outcomes. 
    \item For each outcome, we removed primary studies with an extremely right-skewed outcome distribution. As the variance of the outcome distribution is often incredibly large in such cases, these primary studies receive a weight of nearly 0 in the meta-analysis and cause numerical problems when estimating the between-study variance. As previously considered in the context of meta-analysis of studies reporting medians \cite{mcgrath2019one, mcgrath2020meta, mcgrath2020estimating, ozturk2020meta}, we used Bowley's coefficient \cite{kenney1962keeping} to quantify skewness based on the $S_2$ summary statistics. We removed primary studies when Bowley's coefficient was greater than 0.75. For 20 out of the 24 outcomes meta-analyzed, this criterion removed at most one primary study. For the remaining outcomes, this criterion removed two studies for the neutrophil outcome, four studies for the CRP outcome, six studies for the procalcitonin (PCT) outcome, and removed seven studies for the D-dimer outcome.
    \item We removed the study of Yan et al.\ in the analysis of the D-dimer and CRP outcomes. The extracted sample quantiles for these two outcomes in both groups of patients in this study were around 100-1000 times smaller than the corresponding outcome values in the other primary studies, which caused numerical problems when estimating the between-study variance. 
\end{enumerate}

\subsection{Additional results}

Tables \ref{table:application qe} and \ref{table:application bc} give the pooled difference of means estimates based on the QE and BC approaches, respectively. Tables \ref{table:application tausq} and \ref{table:application I2} gives the estimates of $\tau^2$ and $I^2$, respectively, obtained by all approaches.

\begin{table}[H]
\caption{Pooled difference of means estimates based on the QE approach. \label{table:application qe}}
\centering
\begin{tabular}{@{\extracolsep{6pt}}lll@{}}
  \hline
& \multicolumn{2}{c}{Pooled Difference of Means Estimate [95\% CI]}\\ \cline{2-3}
Outcome & Naïve & Bootstrap \\ 
  \hline
\textbf{Demographics} &  \\ 
\,\,\,\, Age (years) & 13.24 [11.63, 14.85] & 13.22 [11.61, 14.83] \\ 
\textbf{Clinical Values} &  \\ 
\,\,\,\, SpO2-without O2 (\%)  & -7.31 [-9.23, -5.38] & -7.27 [-9.19, -5.34] \\ 
\,\,\,\, Respiratory Rate (per min) & 3.8 [2.83, 4.78] & 3.79 [2.78, 4.79] \\ 
\textbf{Laboratory Values} & \\
\,\,\,\, Hemoglobin (g/L) & -2.22 [-4.92, 0.48] & -2.17 [-4.83, 0.49] \\ 
\,\,\,\, Leukocyte ($10^9$/L) & 3.01 [2.35, 3.66] & 2.97 [2.31, 3.64] \\ 
\,\,\,\, Lymphocyte ($10^9$/L) & -0.37 [-0.43, -0.31] & -0.38 [-0.44, -0.32] \\ 
\,\,\,\, Neutrophil ($10^9$/L) & 3.43 [2.59, 4.27] & 3.42 [2.55, 4.28] \\ 
\,\,\,\, Platelets ($10^9$/L) & -32.77 [-41.73, -23.8] & -31.76 [-41.2, -22.32] \\ 
\,\,\,\, APTT (sec) & 0.82 [-0.47, 2.12] & 0.83 [-0.44, 2.1] \\
\,\,\,\, D-Dimer (mg/L) & 3.02 [1.49, 4.55] & 0.9 [0.39, 1.41] \\  
\,\,\,\, Fibrinogen (g/L) & 0.18 [-0.13, 0.48] & 0.19 [-0.12, 0.5] \\ 
\,\,\,\, INR & 0.1 [0.03, 0.17] & 0.1 [0.03, 0.17] \\ 
\,\,\,\, Prothrombin (sec) & 1.04 [0.8, 1.29] & 1.04 [0.8, 1.28] \\
\,\,\,\, ALAT (U/L) & 4.25 [2.02, 6.48] & 4.11 [1.86, 6.37] \\ 
\,\,\,\, Albumin (g/L) & -4.38 [-5.44, -3.32] & -4.39 [-5.45, -3.33] \\
\,\,\,\, ASAT (U/L) & 16.89 [12.22, 21.56] & 15.83 [11.8, 19.86] \\ 
\,\,\,\, LDH (U/L) & 210.24 [174, 246.48] & 208.34 [171.75, 244.93] \\ 
\,\,\,\, BUN (mmol/L) & 3.72 [2.77, 4.68] & 3.53 [2.63, 4.43] \\ 
\,\,\,\, Creatinine ($\mu$mol/L) & 21.78 [14.37, 29.19] & 20.66 [13.48, 27.85] \\ 
\,\,\,\, CRP (mg/L) & 56.22 [44.53, 67.9] & 54.2 [41.93, 66.48] \\ 
\,\,\,\, IL-6 (pg/mL) & 47.92 [13.66, 82.18] & 4.56 [3.13, 5.99] \\ 
\,\,\,\, PCT (ng/mL) & 0.54 [0.19, 0.89] & 0.29 [0.09, 0.48] \\ 
\,\,\,\, CK (U/L) & 153.79 [86.47, 221.12] & 114.67 [50.63, 178.71] \\
\,\,\,\, CK-MB (U/L) & 5.78 [1, 10.55] & 4.58 [0.19, 8.96] \\ 
   \hline
\end{tabular}
\caption*{\small SpO2 = Oxygen saturation; APTT = activated partial thrombin time; INR = International normalized ratio; ALAT = Alanine transaminase; ASAT = Aspartate transaminase; LDH = Lactate dehydrogenase; BUN = Blood urea nitrogen; CRP = C-reactive
protein; IL-6 = Interleukin-6; PCT = Procalcitonin CK = creatine kinase; CK-MB = creatine kinase–myocardial band.}
\end{table}

\begin{table}[H]
\caption{Pooled difference of means estimates based on the BC approach. \label{table:application bc}}
\centering
\begin{tabular}{@{\extracolsep{6pt}}lll@{}}
  \hline
& \multicolumn{2}{c}{Pooled Difference of Means Estimate [95\% CI]}\\ \cline{2-3}
Outcome & Naïve & Bootstrap \\ 
  \hline
\textbf{Demographics} &  \\ 
\,\,\,\, Age (years) & 13.43 [11.82, 15.04] & 13.43 [11.81, 15.05]  \\ 
\textbf{Clinical Values} &  \\ 
\,\,\,\, SpO2-without O2 (\%)  & -7.41 [-9.43, -5.38] & -7.31 [-9.33, -5.29] \\ 
\,\,\,\, Respiratory Rate (per min) & 3.76 [2.81, 4.72] & 3.73 [2.71, 4.74] \\ 
\textbf{Laboratory Values} & \\
\,\,\,\, Hemoglobin (g/L) & -2.41 [-5.25, 0.43] & -2.29 [-5.03, 0.46]  \\ 
\,\,\,\, Leukocyte ($10^9$/L) & 3.02 [2.37, 3.66] & 3 [2.33, 3.67]  \\ 
\,\,\,\, Lymphocyte ($10^9$/L) & -0.37 [-0.43, -0.31] & -0.38 [-0.44, -0.31]  \\ 
\,\,\,\, Neutrophil ($10^9$/L) & 3.36 [2.51, 4.22] & 3.43 [2.48, 4.38] \\ 
\,\,\,\, Platelets ($10^9$/L) & -35.28 [-45.08, -25.48] & -32.27 [-42.5, -22.03]  \\ 
\,\,\,\, APTT (sec) & 0.82 [-0.47, 2.10] & 0.83 [-0.43, 2.09] \\ 
\,\,\,\, D-Dimer (mg/L) & 3.1 [1.63, 4.56] & 1.18 [0.55, 1.80]  \\ 
\,\,\,\, Fibrinogen (g/L) & 0.16 [-0.17, 0.48] & 0.19 [-0.16, 0.54] \\ 
\,\,\,\, INR & 0.11 [0.03, 0.18] & 0.11 [0.03, 0.18] \\ 
\,\,\,\, Prothrombin (sec) & 1.03 [0.78, 1.28] & 1.02 [0.77, 1.26]  \\ 
\,\,\,\, ALAT (U/L) & 4.47 [2.29, 6.65] & 4.21 [2.05, 6.37]  \\ 
\,\,\,\, Albumin (g/L) & -4.39 [-5.44, -3.35] & -4.41 [-5.47, -3.36]  \\ 
\,\,\,\, ASAT (U/L) & 17.34 [12.43, 22.26] & 16.07 [11.79, 20.36]  \\ 
\,\,\,\, LDH (U/L) & 203.11 [167.74, 238.47] & 202.45 [164.96, 239.94]  \\ 
\,\,\,\, BUN (mmol/L) & 4.31 [2.79, 5.83] & 3.55 [2.65, 4.46]  \\ 
\,\,\,\, Creatinine ($\mu$mol/L) & 21.51 [14.22, 28.8] & 20.37 [13.2, 27.53]  \\ 
\,\,\,\, CRP (mg/L) & 58.82 [47.03, 70.61] & 56.75 [44.33, 69.16]  \\ 
\,\,\,\, IL-6 (pg/mL) & 41.73 [7.82, 75.64] & 4.34 [3.09, 5.6] \\ 
\,\,\,\, PCT (ng/mL) & 0.67 [0.32, 1.02] & 0.45 [0.15, 0.75] \\ 
\,\,\,\, CK (U/L) & 131.09 [72.82, 189.35] & 108.74 [54.72, 162.76]  \\ 
\,\,\,\, CK-MB (U/L) & 6.69 [1.27, 12.1] & 5.24 [0.58, 9.89]  \\ 
   \hline
\end{tabular}
\caption*{\small SpO2 = Oxygen saturation; APTT = activated partial thrombin time; INR = International normalized ratio; ALAT = Alanine transaminase; ASAT = Aspartate transaminase; LDH = Lactate dehydrogenase; BUN = Blood urea nitrogen; CRP = C-reactive
protein; IL-6 = Interleukin-6; PCT = Procalcitonin CK = creatine kinase; CK-MB = creatine kinase–myocardial band.}
\end{table}

\begin{table}[H]
\caption{Estimates of $\tau^2$ obtained by all approaches. \label{table:application tausq}}
\centering
\begin{tabular}{@{\extracolsep{6pt}}lllllll@{}}
  \hline
& \multicolumn{2}{c}{QE Approach} & \multicolumn{2}{c}{BC Approach} & \multicolumn{2}{c}{MLN Approach}\\ \cline{2-3} \cline{4-5} \cline{6-7}
Outcome & Naïve & Boot.\ & Naïve & Boot.\ & Naïve & Boot.\ \\ 
  \hline
  \textbf{Demographics} &  \\ 
\,\,\,\, Age (years) & 27.45 & 26.49 & 27.53 & 25.89 & 28.43 & 27.85 \\ 
\textbf{Clinical Values} &  \\ 
\,\,\,\, SpO2-without O2 (\%)  & 11.99 & 11.66 & 13.38 & 12.74 & 10.59 & 10.35 \\ 
\,\,\,\, Respiratory Rate (per min) & 1.81 & 1.76 & 1.66 & 1.71 & 1.81 & 1.76 \\ 
\textbf{Laboratory Values} & \\
\,\,\,\, Hemoglobin (g/L) & 19.26 & 17.10 & 22.02 & 15.93 & 19.86 & 18.18 \\ 
\,\,\,\, Leukocyte ($10^9$/L) & 2.92 & 2.70 & 2.79 & 2.57 & 2.94 & 2.83 \\ 
\,\,\,\, Lymphocyte ($10^9$/L) & 0.02 & 0.02 & 0.03 & 0.02 & 0.03 & 0.02 \\ 
\,\,\,\, Neutrophil ($10^9$/L) & 3.16 & 2.96 & 3.42 & 3.45 & 3.45 & 3.33 \\ 
\,\,\,\, Platelets ($10^9$/L) & 320.57 & 310.90 & 478.77 & 370.05 & 487.64 & 368.47 \\ 
\,\,\,\, APTT (sec) & 4.31 & 3.91 & 4.18 & 3.51 & 3.89 & 3.37 \\ 
\,\,\,\, D-Dimer (mg/L) & 5.36 & 0.11 & 4.97 & 0.22 & 0.83 & 0.62 \\ 
\,\,\,\, Fibrinogen (g/L) & 0.07 & 0.07 & 0.11 & 0.10 & 0.05 & 0.04 \\ 
\,\,\,\, INR & 0.01 & 0.01 & 0.01 & 0.01 & 0.01 & 0.01 \\ 
\,\,\,\, Prothrombin (sec) & 0.22 & 0.21 & 0.23 & 0.20 & 0.24 & 0.23 \\ 
\,\,\,\, ALAT (U/L) & 11.42 & 2.06 & 13.74 & 0.00 & 14.06 & 9.33 \\ 
\,\,\,\, Albumin (g/L) & 4.34 & 4.23 & 4.19 & 3.96 & 4.38 & 4.30 \\ 
\,\,\,\, ASAT (U/L) & 94.95 & 51.89 & 110.77 & 57.99 & 86.66 & 77.14 \\ 
\,\,\,\, LDH (U/L) & 5269.22 & 5019.80 & 5246.63 & 4952.34 & 5423.54 & 5279.55 \\ 
\,\,\,\, BUN (mmol/L) & 2.74 & 2.11 & 8.18 & 2.00 & 2.01 & 1.90 \\ 
\,\,\,\, Creatinine ($\mu$mol/L) & 282.67 & 247.03 & 275.07 & 235.76 & 258.39 & 247.15 \\ 
\,\,\,\, CRP (mg/L) & 602.62 & 512.44 & 720.26 & 573.54 & 627.39 & 553.02 \\ 
\,\,\,\, IL-6 (pg/mL) & 2437.32 & 1.28 & 2376.03 & 0.69 & 2170.85 & 1.66 \\ 
\,\,\,\, PCT (ng/mL) & 0.14 & 0.03 & 0.20 & 0.09 & 0.09 & 0.10 \\ 
\,\,\,\, CK (U/L) & 10032.66 & 6058.65 & 8264.00 & 4683.80 & 6437.36 & 5422.65 \\ 
\,\,\,\, CK-MB (U/L) & 30.63 & 21.69 & 43.08 & 26.15 & 36.40 & 30.88 \\ 
   \hline
\end{tabular}
\caption*{\small SpO2 = Oxygen saturation; APTT = activated partial thrombin time; INR = International normalized ratio; ALAT = Alanine transaminase; ASAT = Aspartate transaminase; LDH = Lactate dehydrogenase; BUN = Blood urea nitrogen; CRP = C-reactive
protein; IL-6 = Interleukin-6; PCT = Procalcitonin CK = creatine kinase; CK-MB = creatine kinase–myocardial band.}
\end{table}

\begin{table}[H]
\caption{$I^2$ estimates obtained by all approaches. \label{table:application I2}}
\centering
\begin{tabular}{@{\extracolsep{6pt}}lllllll@{}}
  \hline
& \multicolumn{2}{c}{QE Approach} & \multicolumn{2}{c}{BC Approach} & \multicolumn{2}{c}{MLN Approach}\\ \cline{2-3} \cline{4-5} \cline{6-7}
Outcome & Naïve & Boot.\ & Naïve & Boot.\ & Naïve & Boot.\ \\ 
  \hline
  \textbf{Demographics} &  \\ 
\,\,\,\, Age (years) & 88 & 85 & 88 & 82 & 88 & 87 \\ 
\textbf{Clinical Values} &  \\ 
\,\,\,\, SpO2-without O2 (\%)  & 88 & 86 & 89 & 86 & 88 & 86 \\ 
\,\,\,\, Respiratory Rate (per min) & 63 & 57 & 59 & 54 & 65 & 61 \\ 
\textbf{Laboratory Values} & \\
\,\,\,\, Hemoglobin (g/L) & 65 & 59 & 67 & 50 & 67 & 62 \\ 
\,\,\,\, Leukocyte ($10^9$/L) & 84 & 76 & 86 & 73 & 82 & 79 \\ 
\,\,\,\, Lymphocyte ($10^9$/L) & 82 & 72 & 85 & 71 & 90 & 72 \\ 
\,\,\,\, Neutrophil ($10^9$/L) & 90 & 86 & 92 & 86 & 90 & 88 \\ 
\,\,\,\, Platelets ($10^9$/L) & 64 & 57 & 77 & 56 & 72 & 62 \\ 
\,\,\,\, APTT (sec) & 82 & 77 & 81 & 71 & 81 & 76 \\ 
\,\,\,\, D-Dimer (mg/L) & 93 & 11 & 95 & 18 & 82 & 59 \\ 
\,\,\,\, Fibrinogen (g/L) & 52 & 45 & 68 & 50 & 39 & 34 \\ 
\,\,\,\, INR & 85 & 81 & 85 & 78 & 88 & 84 \\ 
\,\,\,\, Prothrombin (sec) & 71 & 65 & 70 & 60 & 74 & 71 \\ 
\,\,\,\, ALAT (U/L) & 33 & 5 & 44 & 0 & 39 & 25 \\ 
\,\,\,\, Albumin (g/L) & 87 & 84 & 86 & 79 & 88 & 86 \\ 
\,\,\,\, ASAT (U/L) & 82 & 62 & 85 & 62 & 81 & 76 \\ 
\,\,\,\, LDH (U/L) & 82 & 78 & 85 & 75 & 82 & 79 \\ 
\,\,\,\, BUN (mmol/L) & 87 & 79 & 95 & 75 & 84 & 80 \\ 
\,\,\,\, Creatinine ($\mu$mol/L) & 87 & 82 & 88 & 78 & 86 & 84 \\ 
\,\,\,\, CRP (mg/L) & 76 & 62 & 86 & 65 & 69 & 59 \\ 
\,\,\,\, IL-6 (pg/mL) & 100 & 34 & 100 & 20 & 100 & 45 \\ 
\,\,\,\, PCT (ng/mL) & 93 & 49 & 95 & 77 & 98 & 91 \\ 
\,\,\,\, CK (U/L) & 86 & 71 & 86 & 66 & 82 & 76 \\ 
\,\,\,\, CK-MB (U/L) & 93 & 88 & 96 & 89 & 95 & 92 \\ 
   \hline
\end{tabular}
\caption*{\small SpO2 = Oxygen saturation; APTT = activated partial thrombin time; INR = International normalized ratio; ALAT = Alanine transaminase; ASAT = Aspartate transaminase; LDH = Lactate dehydrogenase; BUN = Blood urea nitrogen; CRP = C-reactive
protein; IL-6 = Interleukin-6; PCT = Procalcitonin CK = creatine kinase; CK-MB = creatine kinase–myocardial band.}
\end{table}

\end{document}